\documentclass[sigconf]{acmart}
\usepackage[utf8]{inputenc}
\usepackage[english]{babel}
\usepackage{xspace}
\usepackage{todonotes}
\usepackage{tcolorbox}
\usepackage{amsmath,amssymb,amsfonts}
\usepackage[ruled,linesnumbered]{algorithm2e}
\usepackage{algorithmic}
\usepackage{textcomp}
\usepackage{CJKutf8}
\usepackage{bm}
\usepackage{enumitem}
\usepackage{amsmath}
\usepackage{graphics}
\usepackage{epsfig}
\usepackage{xpinyin}

\usepackage{amsfonts}
\usepackage{booktabs}
\usepackage{multirow}
\newcommand{\techname}{TransRepair\xspace}
\newcommand{\jie}[1]{\textcolor{blue}{  #1}}

\AtBeginDocument{%
  \providecommand\BibTeX{{%
    \normalfont B\kern-0.5em{\scshape i\kern-0.25em b}\kern-0.8em\TeX}}}

\setcopyright{acmcopyright}
\copyrightyear{2018}
\acmYear{2018}
\acmDOI{10.1145/1122445.1122456}

\acmConference[]{Preprint}{Under review}{}
\acmPrice{15.00}
\acmISBN{978-1-4503-9999-9/18/06}

\renewcommand\footnotetextcopyrightpermission[1]{}

\begin{document}

\title{Automatic Testing and Improvement of Machine Translation}


\author{Zeyu Sun}
\affiliation{%
  \institution{Peking University}}
\email{szy_@pku.edu.cn}

\author{Jie M. Zhang}
\authornote{Corresponding and co-first author.}
\affiliation{%
 \institution{University College London}
 }
 \email{jie.zhang@ucl.ac.uk}

\author{Mark Harman}
\affiliation{%
 \institution{Facebook London}
 \institution{University College London}}
 \email{mark.harman@ucl.ac.uk}

\author{Mike Papadakis}
\affiliation{%
 \institution{University of Luxembourg}
 }
 \email{mike.papadakis@gmail.com}
 
 \author{Lu Zhang}
\affiliation{%
  \institution{Peking University}}
\email{zhanglucs@pku.edu.cn}

\renewcommand{\shortauthors}{Sun and Zhang, et al.}

\begin{abstract}
This paper presents \techname, a fully automatic approach for testing and repairing the consistency of machine translation systems. 
\techname combines mutation with metamorphic testing to detect inconsistency bugs (without access to human oracles). 
It then adopts probability-reference or cross-reference to post-process the translations, in a grey-box or black-box manner, to repair the inconsistencies. 
Our evaluation on two state-of-the-art translators, Google Translate and Transformer, indicates that \techname has a high precision (99\%) on generating input pairs with consistent translations. 
With these tests, using automatic consistency metrics and manual assessment, we find that Google Translate and Transformer have approximately 36\% and 40\% inconsistency bugs.
Black-box repair fixes 28\% and 19\% bugs on average for Google Translate and Transformer.
Grey-box repair fixes 30\% bugs on average for Transformer.
Manual inspection indicates that the translations repaired by our approach improve consistency in 87\% of cases (degrading it in 2\%), and that our repairs have better translation acceptability in 27\% of the cases (worse in 8\%). 
\end{abstract}





\maketitle
\section{Introduction}

Machine learning has been successful in providing general-purpose natural language translation systems, with many systems able to translate between thousands of pairs of languages effectively in real time \cite{hazelwoodetal:fbleaner18}.
Nevertheless, such translation systems are not perfect and the bugs that users experience have a different character from those on traditional, non-machine learning-based, software systems ~\cite{zhang2019machine,belinkov2017synthetic,khayrallah2018impact,karpukhin2019training}.

The consequences of mistranslation have long been studied and  their effects have been shown to be serious.
For example, the infamous historic mistranslation of Article 17 of the Treaty of Uccialli reportedly led to war  \cite{giglio:uccialli}.
Such truly profound and far-reaching consequences of mistranslation are also reportedly becoming a serious and potent source of international tension and conflict \cite{mason:lost}.

\begin{figure}[h]\small
    \center
            \includegraphics[scale=0.28] {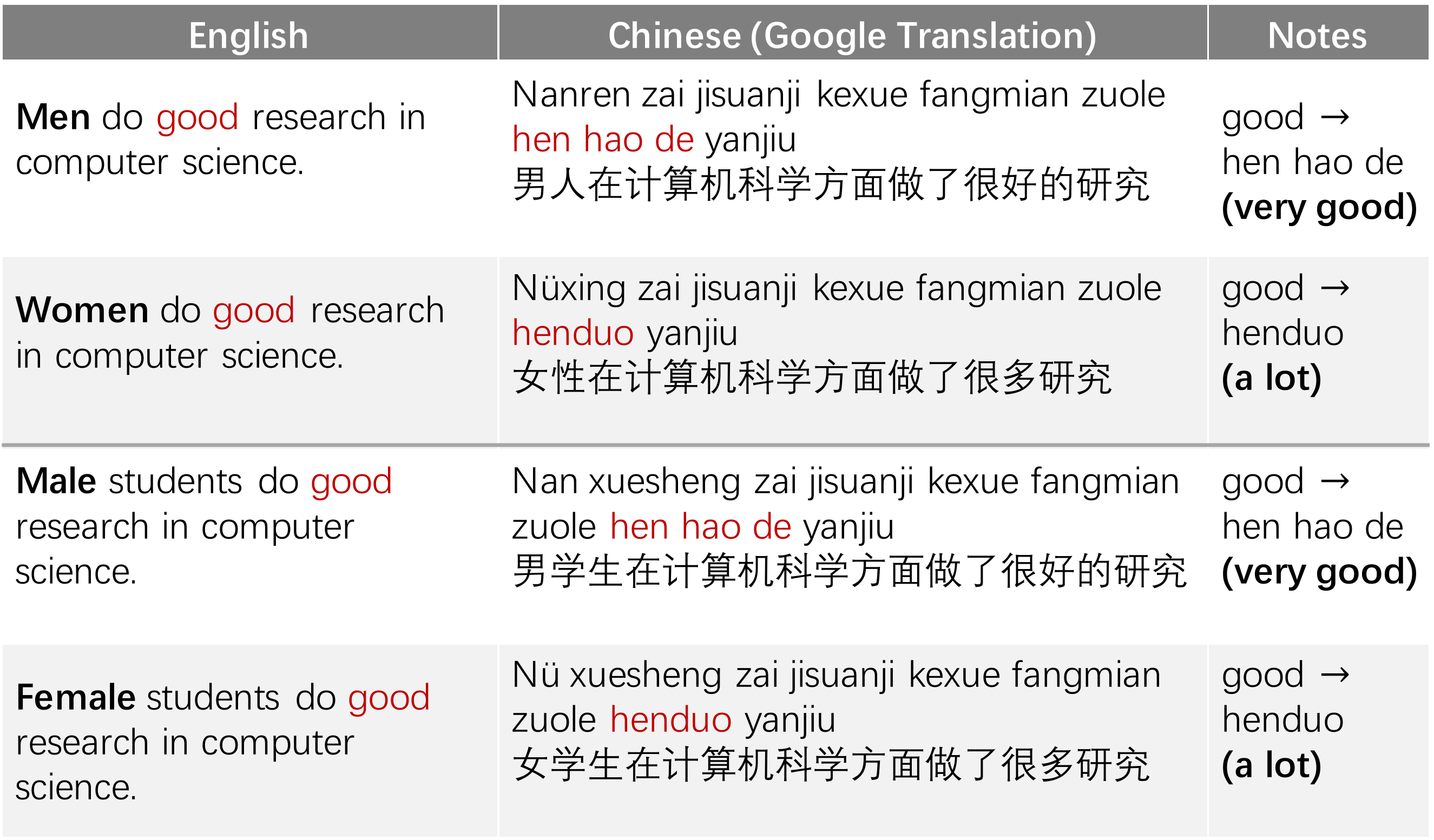}
                        \vspace{-2mm}
    \caption{Examples of fairness issues brought by translation inconsistency (from Google Translate)}
    \label{fig:motivatingexample}
\end{figure}

The consequences of mistranslation through machine-based translators have also been shown to be serious.
For example, machine translations have  been shown to exhibit pernicious fairness bugs that disproportionately harm specific user constituencies~\cite{sexisttranslator}.
We have found such examples of fairness bugs in widely used industrial strength translation systems. 
Figure~\ref{fig:motivatingexample} shows several such Google Translate results for the language pair (English $\rightarrow$ Chinese)\footnote{The four translations were obtained on 23rd July, 2019. 
These examples are purely for illustration purposes, and are not intended as a criticism of Google Translate. It is likely that other mainstream translation technologies will have similar issues.}.
As can be seen from the figure, Google Translate translates `good' into `hen hao de' (which means `very good') when the subject is `men' or `male students'.
However, interestingly, but also sadly, it translates `good' into `hen duo' (which means `a lot') when the subject is `women' or `female students'\footnote{Similar issues also exist in translations between other languages. With a cursory check, we already found a case with German$\rightarrow$Chinese.}. 

Such inconsistency may confuse users, and is also clearly \emph{unfair} to female researchers in computer science; producing `a lot' of research is clearly a more pejorative interpretation, when compared to producing `very good' research.
To avoid such unfair translations (at scale), we need techniques that can automatically identify and remedy such inconsistencies.

To tackle this problem, we introduce a combined testing and repair approach that automatically generates tests for real-world machine translation systems, and  automatically repairs the mistranslations found in the testing phase.
As we show in this paper, we need to rethink the conventional approaches to both testing and repair in order to apply them to natural language translation.

Existing work has tested whether machine translation systems provide stable translations for semantically-equal transformations, such as synonym replacement (e.g., buy$\rightarrow$purchase)~\cite{cheng2018towards} or abbreviation replacement (e.g., what's$\rightarrow$what is)~\cite{ribeiro2018semantically}.
However, no previous work has focused on the testing and repair of translation inconsistency regarding context-similar transformation;
the transformation between sentences that have similar word embeddings~\cite{goldberg2014word2vec} yet share context in the corpus (e.g., simple gender-based transformations, such as  boys$\rightarrow$girls).

In order to tackle the testing problem, 
we introduce an approach that combines mutation \cite{yjmh:analysis} with metamorphic testing \cite{chen1998metamorphic,zhang2014search}. 
The approach conducts context-similar mutation to generate mutated sentences that can be used as test inputs for the translator under test.
When a context-similar mutation yields above-threshold disruption to the translation of the non-mutated part, the approach reports an inconsistency bug.



Traditional approaches to `repairing' machine learning systems typically use data augmentation or algorithm optimisation.
These approaches can best be characterised as to ``improve'' the overall effectiveness of the machine learner, rather than specific repairs for individual bugs;
they also need data collection/labelling and model retraining, which usually have a high cost.


Traditional approaches to `repairing' software bugs are white box, because the techniques need to identify the line(s) of source code that need(s) to be modified in order to implement a fix.
Such approaches inherently cannot be applied to fix software for which source code is unavailable, such as third-party code.

Our insight is that by combining the results of repeated (and potentially inconsistent) output from a system, we can implement a light-weight {\em black-box} repair technique as a kind of `post-processing' phase that targets specific bugs.
Our approach is the first repair technique to repair a system in a purely black-box manner.
We believe that black-box repair has considerable potential benefits, beyond the specific application of machine translation repair. 
It is the only available approach when the software engineer is presented with bugs in systems for which no source code is available.

We demonstrate not only that black-box repair is feasible, but that it can scale to real world industrial-strength translation systems, such as Google Translate.
We also present results for grey-box repair for which the predictive probability is available.

\techname is evaluated on two state-of-the-art machine translation systems, Google Translate and Transformer~\cite{vaswani2017attention}.
In particular, we focus on the translation between the top-two most widely-spoken languages: English and Chinese. 
These languages each have over one  billion speakers worldwide~\cite{eberhard2019ethnologue}.
Nevertheless, only 10 million people in China (less than 1\% of the population) are able to communicate via English~\cite{chinese01,wei2012statistics}.
Since so few people are able to speak both languages, machine translation is often attractive and sometimes necessary and unavoidable.

Our results indicate that \techname generates valid test inputs effectively with a precision of 99\%; 2) \techname automatically reports inconsistency bugs effectively with the learnt thresholds, with a mean F-measure of 0.82/0.88 for Google Translate/Transformer;
3) Both Google Translate and Transformer have inconsistency bugs. 
Automated consistency metrics and manual inspection reveal that Google Translate has approximately 36\% inconsistent translations on our generated test inputs. 
4) Black-box repair reduces 28\% and 19\% of the bugs of Google Translate and Transformer.
Grey-box reduces 30\% of the Transformer bugs.
Manual inspection indicates that the repaired translations improve consistency in 87\% of the cases (reducing it in only 2\%), and have better translation acceptability\footnote{We use ``acceptability'' to capture the property that a translation meets human assessment of a reasonable (aka acceptable) translation} in 27\% of the cases (worse in only 8\%)

\section{Approach}

This section introduces the overview and the details of each step for \techname.

\subsection{Overview}
    
\begin{figure*}
    	    \centering
      		\includegraphics[width=0.8\linewidth]{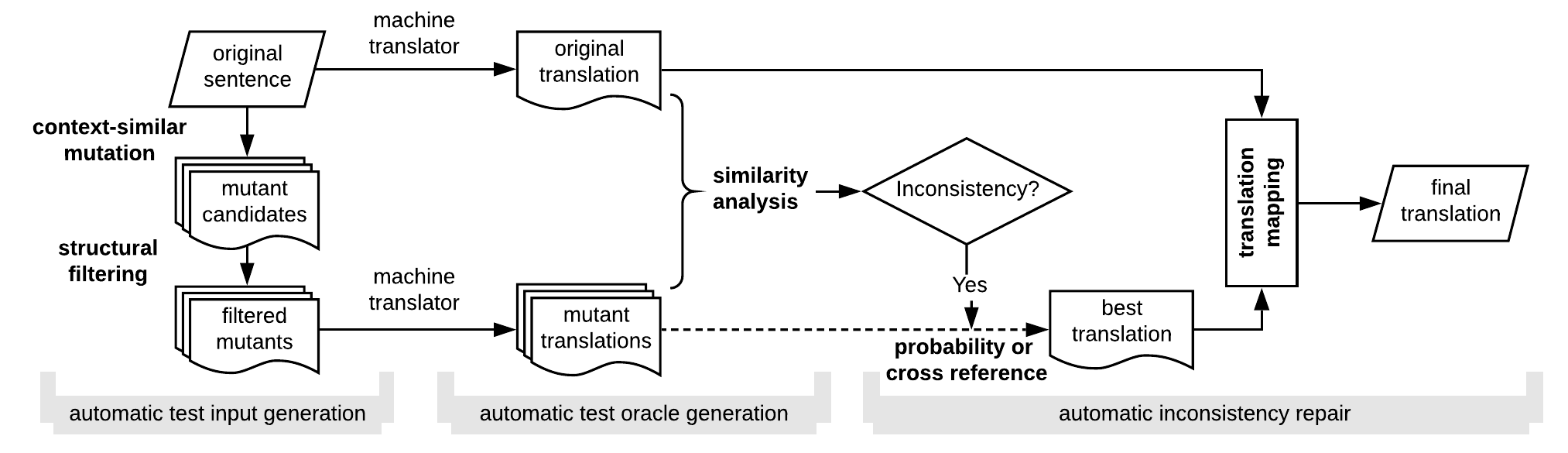}
      		\vspace{-4mm}
    		\caption{Overview of how \techname tests and repairs machine translation inconsistencies.}
    		\label{Overview}
\end{figure*}

A high level view of \techname is presented in Figure~\ref{Overview}. From this Figure it can be seen that  
\techname automatically tests and repairs the inconsistency of machine translation based on the following three major steps: 

\textbf{1) Automatic test input generation.} This step generates transformed sentences (test inputs)  to be used for consistency testing. 
For each sentence, \techname conducts sentence mutations via context-similar word replacement.  
The generated mutant candidates are filtered using a grammar check. 
The mutants that pass the grammar check are then regarded as the final test inputs for the machine translator under test. Details are presented in Section~\ref{sec:testinputgeneration}.

\textbf{2) Automatic test oracle generation.} This step introduces the generation of oracles, which are used for identifying inconsistent translations (bugs). 
In this step, we rely on the metamorphic relationship between translation inputs and translation outputs. 
The idea is that translation outputs from both the original sentence and its context-similar mutant(s) should have a certain degree of consistency modulo the mutated word. 
We use similarity metrics that measure the degree of consistency between translated outputs as test oracles.   
Details of this step are presented in Section~\ref{sec:testoracleidentification}. We explore four similarity metrics, which are described in Section~\ref{sec:similaritymetrics}.

\textbf{3) Automatic inconsistency repair.} 
This step automatically repairs the inconsistent translation. 
\techname applies black-box and grey-box approaches, which transform the original translation based on the best translation among the mutants.
We explore two ways of choosing the best translation, one using predictive probability, the other using cross-reference.
Details of this step are given in Section~\ref{sec:repair}.

\subsection{Automatic Test Input Generation}
\label{sec:testinputgeneration}
The input generation process contains the following steps.


\subsubsection{Context-similarity Corpus Building}

To conduct context-similar word replacement, the key step is to find a word(s) that can be replaced with other(s) (similar ones) without hurting the sentence structure. The new sentence generated via word replacement should yield consistent translations with the original.

Word vectors capture the meaning of a word through their context~\cite{pennington2014glove}. 
To measure the similarity, we use word vectors trained from text corpora. 
In our approach, the word similarity between two words $w_1$ and $w_2$, denoted by $\mathrm{sim}(w_1, w_2)$, is computed by the formula below, where $\bm v_{x}$ denotes the vector of the word $x$. 
        \begin{equation}
            \mathrm{sim}(w_1, w_2) = \frac{\bm v_{w_1} \bm v_{w_2}}{|\bm v_{w_1}||\bm v_{w_2}|}
            \label{eq:cos_sim}
        \end{equation}
        
To construct a reliable context-similar corpus, we take two word-vector models and use the intersection of their trained results.
The first model is GloVe~\cite{pennington2014glove}, which is trained from Wikipedia 2014 data~\cite{wikipedia} and GigaWord 5~\cite{gigaword}. 
The second model is SpaCy~\cite{SpaCy}, which is a multi-task CNN trained on OntoNotes~\cite{OntoNotes} including the data collected from telephone conversations, newswire, newsgroups, broadcast news, broadcast conversation, and weblogs. 
When two words have a similarity of over 0.9 for both models, we deem the word pair to be context-similar and place it in the context-similarity corpus.
In total, we collected 131,933 word pairs using this approach.

\subsubsection{Translation Input Mutation}
\label{sec:translationinputmutation}
We introduce word replacement and structural filtering respectively in the following.

\paragraph{Word replacement}
For each word in the original sentence, we search to determine whether there is a match in our corpus.
If we find a match, we replace the word with its context-similar one and generate the resulting mutated input sentence.
Compared with the original sentence, each mutated sentence contains a single replaced word. 
To reduce the possibility of generating mutants that do not parse, we only replace nouns, adjectives, and numbers.

\paragraph{Structural filtering}
The generated mutated sentence may fail to parse, because the replaced word may not fit the context of the new sentence.
For example, ``one'' and ``another'' are context-similar words, but ``a good one'' parses, while ``a good another'' does not. 
To address such parsing failures, we apply additional constraints to sanity check the generated mutants. 
In particular, we apply structural filtering, based on the Stanford Parser~\cite{manningEtAl2014}.
Suppose the original sentence is $s=w_1,w_2,...,w_i,...,w_n$, 
the mutated sentence is $s'=w_1,w_2,...,w_i',...,w_n$, where $w_i$ in $s$ is replaced with $w_i'$ in $s'$.
For each sentence, the Stanford Parser outputs $l(w_i)$, the part-of-speech tag of each word used in the Penn Treebank Project~\cite{taylor2003penn}.
If $l(w_i)\neq l(w_i')$, we remove $s'$ from the mutant candidates because the mutation yields changes in the syntactic structure.

We manually inspect the quality of the generated inputs and report results in Section~\ref{sec:results}.

\subsection{Automatic Test Oracle Generation}
\label{sec:testoracleidentification}

To perform testing we need to augment the test inputs we generate with test oracles, i.e., predicates that check whether an inconsistency bug has been found. 
To do so, we assume that the unchanged parts of the
sentences preserve their adequacy and fluency modulo the mutated word.
Adequacy means whether the translation conveys identical meaning, whether there is information lost, added, or distorted;
Fluency means whether the output is fluent and grammatically correct~\cite{doddington2002automatic,graham2012measurement}.

Let $t(s)$ and $t(s')$
be the translations of sentences $s$ and $s'$ that were produced by
replacing the word $w$ (in $s$) with $w'$ (in $s'$). 
Thus, we would like to check the similarity between
$t(s)$ and $t(s')$ when ignoring the translations for
words $w$ and $w'$. 
Unfortunately, it is not easy to strip the effect of $w$ and $w'$, because a) $w$ and $w'$ may change the entire translation of the sentences, and b) it is not easy to accurately map the words $w$ and $w'$ with their respective one(s) in the translated text.

To bypass this problem, we calculate the similarity of subsequences of $t(s)$ and $t(s')$, and use the largest similarity to approximate the consistency level between $t(s)$ and $t(s')$.
Algorithm~\ref{alg:sim} shows the process.
For $t(s)$ and $t(s')$, 
we first apply GNU Wdiff~\cite{wdiff} to get the difference slices (Line 1). 
GNU Wdiff compares sentences on word basis, and is useful for comparing two texts in which a few words have been changed~\cite{wdiff}.
With Wdiff, the difference slices of two sentences ``\textbf{\textcolor{blue}{A}} B C \textbf{\textcolor{blue}{D}} F'' and ``\textbf{\textcolor{blue}{B}} B C \textbf{\textcolor{blue}{G H}} F" are represented as ``A'', ``D'' and ``B'', ``G H'' for the two sentences, respectively. 

The diff slices of $t(s)$ and $t(s')$ are saved to set $B_s$ and $B_{s'}$\footnote{Long slices are unlikely to correspond to the mutated word, we thus only keep slices that are no longer than 5 words.}.
We then delete a slice from the translations, one at a time (Line 5 and Line 9).
Each slice corresponds to one subsequence with this slice deleted. 
For the example above,  ``\textbf{\textcolor{blue}{A}} B C \textbf{\textcolor{blue}{D}} F'' will have two subsequences: ``B C D F'' (deleting ``A'') and ``A B C F'' (deleting ``D'').
These new subsequences of $t(s)$/$t(s')$ are added into set $T_o$/$T_m$ (Line 6 and Line 10).

For each element in set $T_o$, we compute its similarity\footnote{We explore four types of similarity metrics in this paper (see full details in Section~\ref{sec:similaritymetrics}).} with each element in the set $T_m$ (Line 15). 
Thus, we get $|T_o|*|T_m|$ similarity scores, where $|T_o|$ and $|T_m|$ is the size of $T_o$ and $T_m$. 
We then use the highest similarity as the result of the final consistency score (Lines 16). 

This configuration reduces the influence of the mutated word, and helps to select an inconsistency upper bound. 
Even if the two subsequences with the largest similarity contain the replaced word, other sentence parts have worse similarity, so it is unlikely that this case is biased by the replaced word (leads to false positives).

Details about the experimental setup and the results of the threshold setup are discussed in Section~\ref{sec:rq2}. Additionally, we evaluate the effectiveness of our consistency measurements via manual inspection. These results are presented in Section~\ref{sec:metrichumancorr}.

\begin{algorithm}\small
    \caption{Process of obtaining consistency score \label{alg:sim}} 
    \KwData{$t(s)$: translation of the original sentence; $t(s')$: translation of the mutant }
 \KwResult{ConScore: Consistency score between $t(s)$ and $t(s')$ }
    $B_s, B_{s'} = \mathrm{Wdiff}(t(s), t(s'))$\\
    $T_o = \{ t(s) \}$\\
    $T_m = \{ t(s') \}$\\
    \For{each subsequence $b_s \in B_s$} 
        { 
                $r = \mathrm{DeleteSub}(t(s), b_s)$\\
                $T_o = T_o \cup \{r\}$\\
            }
    \For{each subsequence $b_{s'} \in B_{s'}$}
    {
              $r' = \mathrm{DeleteSub}(t(s'), b_{s'})$\\
              $T_m = T_m \cup \{r'\}$\\
          }
    ConScore = -1\\
    \For{each sentence $a \in T_o$}
    {
        \For{each sentence $b \in T_m$}
        {
             Sim = $\mathrm{ComputeSimilarity}(a, b)$\\
            ConScore = $\mathrm{Max}(ConScore, Sim)$\\
            }
        }
    \Return ConScore
\end{algorithm}

\subsection{Automatic Inconsistency Repair}
\label{sec:repair}

We first introduce the overall repair process, then introduce two mutant translation ranking approaches (probability and cross-reference).

\subsubsection{Overall Repair Process}

\begin{algorithm}\small
    \caption{Process of automatic repair \label{alg:rep}} 
    \KwData{
    $s$: a sentence input; 
    $t(s)$: translation of $s$; 
    $s_1, s_2, ..., s_n$: mutants of $s$;
    $t(s_1), t(s_2), ..., t(s_n)$: translations of the mutants;}
 \KwResult{NewTrans: repaired translation for $s$}
    $T = \{ (s, t(s)), (s_1, t(s_1)), (s_2, t(s_2)), ..., (s_n, t(s_n)) \}$\\
    $\mathrm{OrderedList} = \mathrm{Rank}(T)$\\
    $a(s) = \mathrm{wordAlignment}(s, t(s))$\\
    $\mathrm{NewTrans} = t(s)$\\
    \For{each sentence and its translation $s_r, t(s_r)$ $\in$ $\mathrm{OrderedList}$}
    { 
        \If{$s_r == s$}
            {$\mathrm{break}$\\}
        $a(s_r) = \mathrm{wordAlignment}(s_r, t(s_r))$ \\
        $w_i, w_i^r = \mathrm{getReplacedWord}(s, s_r)$\\
        $t(w_i) = \mathrm{getTranslatedWord}(w_i, a(s))$\\
        $t(w_i^r) = \mathrm{getTranslatedWord}(w_i^r, a(s_r))$\\
        \If{$\mathrm{isnumeric}(w_i) != \mathrm{isnumeric}(t(w_i)) $ \textbf{or} $ \mathrm{isnumeric}(w_i^r) != \mathrm{isnumeric}(t(w_i^r))$}
        {
            $\mathrm{continue}$\\
        }
        $t^r(s_r) = \mathrm{mapBack(t(s), t(s_r), s, s_r, a(s), a(s_r))}$\\
        \If{\textbf{not} ( $\mathrm{isnumeric}(w_i)$ \textbf{and} $\mathrm{isnumeric}(w_i^r)$)}
        {
            \If{$\mathrm{structure}(t^r(s_r)) != \mathrm{structure}(t(s_r))$}
            {
                $\mathrm{continue}$\\
            }
        }
        \If{$\mathrm{isTest}(s)$}{
            $s_o, t(s_o) = \mathrm{getRepairedResult}(s)$\\
            \If{not $isConsistent(t(s_o), t^r(s_r))$}
            {
                $\mathrm{continue}$\\
            }
        }
       NewTrans = $t^r(s_r)$\\
        $\mathrm{break}$\\  
    }
    \Return NewTrans
\end{algorithm}

First, we repair the translation of the original sentences and then we seek to find a translation for the mutant, which must pass our consistency test.

Algorithm~\ref{alg:rep} shows the repair process.
For $t(s)$ which has been revealed to have inconsistency bug(s), 
we generate a set of mutants and get their translations $t(s_1), t(s_2), ..., t(s_n)$.
These mutants and their translations, together with the original sentence and its translation, are put into a dictionary, $T$ (Line 1).
We then rank the elements in $T$, in descending order, using the predictive probability or cross-reference, and put the results in $OrderedList$ (Line 2).
The details of probability and cross-reference ranking are given in Section~\ref{sec:probabilitydetails} and Section~\ref{sec:crossreferencedetails}.

Next, we apply word alignment to obtain the mapped words between $s$ and $t(s)$ as $a(s)$ (Line 3). 
Word alignment is a natural language processing technique that connects two words if and only if they have a translation relationship.
In particular, we adopt the technique proposed by Liu et al.~\cite{liu2015contrastive}, which uses a latent-variable log-linear model for unsupervised word alignment.
We then check whether a sentence pair $(s_r,t(s_r))$ in $OrderedList$ can be adopted to repair the original translation.
We follow the ranking order, until we find one mutant translation that is acceptable for inconsistency repair.

If $s_r$ is the original sentence ($s_r==s$), it means the original translation is deemed a better choice than other mutant translations and so we will not touch it 
(Lines 6-8). 
Otherwise, we do the same alignment to $s_1$ and $t(s_1)$ as to $s$ and $t(s)$. The variables $w_i$, $w_i^r$ denote the replaced words in $s$, $s_r$ and we get the translated words $t(w_i)$, $t(w_i^r)$ through the alignment (Lines 9-12).

Word alignment is not 100\% accurate. If we directly map the translation by replacing $t(w_i^r)$ with $t(w_i)$, it may lead to grammatical errors or context mismatches. 
We adopt the following strategies to judge whether the replacement is acceptable. 
1) We constrain that $w_i$, $w_i^r$ and $t(w_i)$, $t(w_i^r)$ must belong to the same type (i.e., numeric or non-numeric) (Line 13-15).
2) If the replaced words are of the non-numeric type, we apply Stanford Parser to check whether the replacement would lead to structural changes (Line 17-21). 

When repairing the translation of the mutated input (Line 22), we get the repaired result of the original sentence (Line 23), then check whether the translation solution candidate is consistent with the repaired translation of the original sentence (Line 24-26).
If not, we proceed by checking other repair candidates. 
    
\subsubsection{Translation Ranking based on Probability}
\label{sec:probabilitydetails}

For a sentence $s$ and its mutants $S={s_1,s_2, ...s_n}$,
let $t(s)$ be the translation of $s$, 
let $t(s_i)$ be the translation of mutant $s_i$.
This approach records the translation probability for each $t(s_i)$, 
and chooses the mutant with the highest probability as a translation mapping candidate.
The translation of the corresponding mutant will then be mapped back to generate the final repaired translation for $s$ using word alignment.

This is a grey-box repair approach.
It requires neither the training data nor the source code of the training algorithm, but needs the predictive probability provided by the machine translator. 
We call this grey-box because implementors may regard this probability information as an internal attribute of the approach, 
not normally intended to be available to end users. 

\subsubsection{Translation Ranking based on Cross-reference}
\label{sec:crossreferencedetails}

For a sentence $s$ and its mutants $S={s_1,s_2, ...s_n}$,
let $t(s)$ be the translation of $s$, 
and let $t(s_i)$ be the translation of mutant $s_i$.
This approach calculates the similarity among $t(s),t(s_1),t(s_2), ...t(s_n)$,
and uses the translation that maps the best (with the largest mean similarity score) with other translations to map back and repair the previous translation.

This is a black-box repair approach.
It requires only the ability to execute the translator under test and the translation outputs.

\section{Experimental Setup}
\label{sec:experimentalsetup}

In this section, we introduce the experimental setup that evaluates the test input generation, translation inconsistency detection, and translation inconsistency repair.

\subsection{Research questions}
    \noindent

We start our study by assessing the ability of \techname to generate valid and consistent test inputs that can be adopted for consistency testing. Hence we ask:

    

\begin{description}
    \item[RQ1:] \textbf{How accurate are the test inputs of \techname?}
 \end{description}
    

We answer this question by randomly sampling some candidate pairs and checking (manually) whether they are valid. The answer to this question ensures that, \techname, indeed generates inputs that are suitable for consistency checking. 

Given that we found evidence that \techname generates effective test pairs, we turn our attention to the question of how effective these pairs are at detecting consistency bugs. Therefore we ask:

\begin{description}
    \item[RQ2:]\textbf{What is the bug-revealing ability of \techname?}
\end{description}

To answer RQ2 we calculate consistency scores based on similarity metrics to act as test oracles (that determine whether a bug has been detected). To assess the bug-revealing ability of the \techname, we manually check a sample of translations and compare the resulting manual inspection results with automated test results. 
    
Having experimented with fault revelation, we evaluate the repair ability of \techname to see how well it repairs inconsistency bugs. Thus, we ask:
    
\begin{description}
\item[RQ3:] \textbf{What is the bug-repair ability of \techname?}
\end{description}

To answer this question, we record how many inconsistency bugs are repaired (assessed by consistency metrics and manual inspection). 
We also manually examine the translations repaired by \techname, and check whether they improve translation consistency as well as quality.

\subsection{Consistency Metrics}
\label{sec:similaritymetrics}

We explore four widely-adopted similarity metrics for measuring inconsistency.
For ease of illustration, we use $t_1$ to denote the translation output of the original translation input;
we use $t_2$ to denote the translation output of the mutated translation input.

\paragraph{\textbf{LCS-based metric}} It measures the similarity via normalised length of a longest common subsequence between $t_1$ and $t_2$:
\begin{equation}
   M_{LCS} = \frac{len(LCS(t_1,t_2))}{Max(len(t_1),len(t_2))}
\end{equation}
In this formula, $LCS$ is a function that calculates a longest common subsequence~\cite{hunt1977fast} between $t_1$ and $t_2$ that appear in the same relative order.
For example, an LCS for input Sequences ``ABCDGH'' and ``AEDFHR'' is ``ADH'' with a length of 3.

\paragraph{\textbf{ED-based metric}} This metric is based on the edit distance between $t_1$ and $t_2$. 
Edit distance is a way of quantifying how dissimilar two strings are by counting the minimum number of operations required to transform one string into the other~\cite{ristad1998learning}. 
To normalise the edit distance, we use the following formula which has also been adopted in previous work~\cite{gu2018search,zhang2018guiding}.
\begin{equation}
   M_{ED} = 1 - \frac{ED(t_1,t_2)}{Max(len(t_1),len(t_2))}
\end{equation}
In this formula, $ED$ is a function that calculates the edit distance between $t_1$ and $t_2$.

\paragraph{\textbf{tf-idf-based metric}}
tf-idf (term frequency–inverse document frequency) can be used to measure similarity in terms of word frequency.
Each word $w$ has a weighting factor, which is calculated based on the following formula, where $C$ is a text corpus (in this paper we use the training data of Transformer), $|C|$ is the sentence number in $C$, $f_{w}$ is the number of sentences that contain $w$.
\begin{equation}
    w_{idf} = \log((|C| + 1)/(f_{w} + 1))
\end{equation}
We then represent each sentence with the bag-of-words model~\cite{Zhang2010}, which is a simplified representation commonly used in natural language processing. 
In this model, the grammar and the word order is disregarded, only keeping multiplicity (i.e., ``A B C A'' is represented as {``A'':2, ``B'':1, ``C'':1}, namely [2, 1, 1] in vector). 
Each dimension of the vector is multiplied with its weight $w_{idf}$.
We calculate the cosine similarity (Equation~\ref{eq:cos_sim}) of the weighted vectors of $t_1$ and $t_2$ as their final tf-idf-based consistency score.

\paragraph{\textbf{BLEU-based metric}}
The BLEU (BiLingual Evaluation Understudy) is an algorithm that automatically evaluates machine translation quality via checking the correspondence between a machine's output and that of a human.
It can also be adopted to compute the similarity between the translation of the original sentence and the translation of a mutant. 
Details, description, and motivation for the BLEU score can be found in the translation literature~\cite{papineni2002bleu}. Due to lack of space we only provide an overview here.

BLEU first counts the number of matched subsequences between sentences and
computes a precision $p_n$ (which is called modified n-gram precision~\cite{papineni2002bleu}, where $n$ means the subsequence length). 
For example, in sentences ``A A B C'' ($s_1$) and  ``A B B C'' ($s_2$) there are three 2-gram subsequences in $s_2$: $AB$, $BB$, and $BC$.
Two of them are matched with those from $s_1$: $AB$ and $BC$.
Thus, $p_2$ is 2/3.

As well as $p_n$, the calculation of BLEU score also requires an exponential brevity penalty factor $BP$ (to penalise overaly short translations), which is shown by Formula~\ref{form6}. 
$c$ denotes the length of $t(s_i)$ and $r$ is the length of $t(s)$.
\begin{equation}
\label{form6}
    \mathrm{BP}=\left\{\begin{array}{ll}{1} & {\text { if } c>r} \\ {e^{(1-r / c)}} & {\text { if } c \leq r}\end{array}\right.
\end{equation}
The BLEU score is finally computed by Formula~\ref{form7}, where $w_n=\frac{1}{N}$ (we use $N=4$ in this paper) is the uniform weights.
\begin{equation}
\label{form7}
    \mathrm{BLEU}=\mathrm{BP} \cdot \exp \left(\sum_{n=1}^{N} w_{n} \log p_{n}\right).
\end{equation}
Since BLEU is unidirectional (i.e., $BLEU(s,s')\neq BLEU(s',s)$), we use the higher score for measuring the similarity between $s$ and $s'$.
This is consistent with our intention in Algorithm~\ref{alg:sim}: to get an upper bound of the consistency, thereby avoiding false positive claims about translation bugs. 

        
\subsection{Machine Translators}
Our experiment considers both industrial and state-of-the-art research-oriented machine translators. 
One is Google Translate~\cite{googletranslate} (abbreviated as GT in the results section), a widely used machine translation service developed by Google. 
The other is Transformer~\cite{vaswani2017attention}, a translator studied by the research community.

We use Google translate, because it is an example of a system that forces us to perform black-box repairs; we have no access to the training data nor the code of the translation system, and therefore any improvements, by definition, can only have been achieved by a black-box approach. Also, of course, it is a production-quality mainstream translation system, making the results more interesting.

We use the default setup to train Transformer.
Transformer is trained based on three datasets:
the CWMT dataset~\cite{CWMT} with 7,086,820 parallel sentences, 
the UN dataset~\cite{UN} with 15,886,041 parallel sentences, 
and the News Commentary dataset~\cite{news} with 252,777 parallel sentences as the training data.
The validation data (to help tune hyper-parameters) is also from the News Commentary and contains 2,002 parallel sentences.
Transformer runs with the Tensor2Tensor deep learning library~\cite{tensor2tensor}.
To get a more effective translator, we train the model for 500,000 epochs.

\subsection{Test Set } 
\label{sec:datasets}

Following previous machine translation research
~\cite{DBLP:journals/corr/abs-1803-05567,hao-etal-2019-modeling}, we use a test set from the News Commentary~\cite{news} dataset for both Google Translate and Transformer.
The test set contains 2,001 parallel sentences and are different from the training set and validation set. 
The Chinese sentences in our experiments are in the form of characters. set\footnote{The Chinese sentences in our experiments are in the form of characters.}.

\vspace{3mm}
Our experiments were conducted on Ubuntu $16.04$ with 256GB RAM and four Intel E5-2620 v4 CPUs ($2.10$ GHz), which contains 32 cores all together.
The neural networks we used were all trained on a single Nvidia Titan RTX (24 GB memory).

\section{Results}
\label{sec:results}

This section reports the results that answer our research questions.

\subsection{Effectiveness on Input Generation (RQ1)}
We start by answering RQ1.  
For each test sentence, we generate mutants and check whether they pass the structural filtering (see more in Section~\ref{sec:translationinputmutation}). 
In particular, for each sentence we generate up to 5 mutants that pass through the filter (we study the influence of the number of mutants in Section~\ref{sec:mutantnumberinfluence}).
For the 2,001 test sentences, 21,960 mutant candidates are generated, with 17,268 discarded by structural filtering.
In the rest of our experiment we use the remaining 4,692 mutants, which are paired with 1,323 sentences, as test inputs.

To manually assess whether these test inputs are qualified for detecting translation inconsistency, we randomly sampled 400 of them.
The first two authors manually checked the validity of the inputs, i.e., whether the replaced word in the mutant leads to grammatical errors and whether the mutant ought to have consistent translations with the original sentence.
This validation step reveals three invalid mutants: 
1) \emph{He was a kind spirit with a big heart}: kind $\rightarrow$ sort; 2) \emph{Two earthquakes with magnitude 4.4 and 4.5 respectively}: Two $\rightarrow$ Six;
3) \emph{It is in itself a great shame}: great $\rightarrow$ good.

The remaining 397 of the 400 meet the two validity criteria, indicating a precision of 99\%.
We conclude that our two strategies for the intersection of two word2vec models, and the use of Stanford Parser as a filter have a high probability of yielding valid test sentences.
The 400 mutants and the manual assessment results can be found on the \techname homepage~\cite{homepage}.

In the next section we use the 4,692 mutants (from the 1,323 original sentences) to examine the test translation consistency of machine translation systems.

\begin{tcolorbox}
Answer to \textbf{RQ1}: \techname has good precision (99\%) for generating test sentences that are grammatically correct and yield consistent translations.
\end{tcolorbox}
	   
\subsection{Inconsistency-revealing Ability of \techname (RQ2)}
\label{sec:rq2}

This section answers RQ2, i.e., investigates the inconsistency-revealing ability of \techname. 
To answer this question, we investigate: 1) the consistency metric values between the mutant and the original translations; 2) the manual inspection results of translation inconsistency.
We also explore how close the consistency metrics and manual inspection are in evaluating inconsistency.

\paragraph{Consistency Metric Values}
We translate the 4,692 generated inputs with Google Translate and Transformer, and compare them with the translations of the original sentences, following the steps of Algorithm~\ref{alg:sim}.
For each mutant, we calculate four consistency scores, each one corresponding to one of the similarity metrics (outlined in Section~\ref{sec:similaritymetrics}).

\begin{figure}
    	    \centering
      		\includegraphics[width=0.95\linewidth]{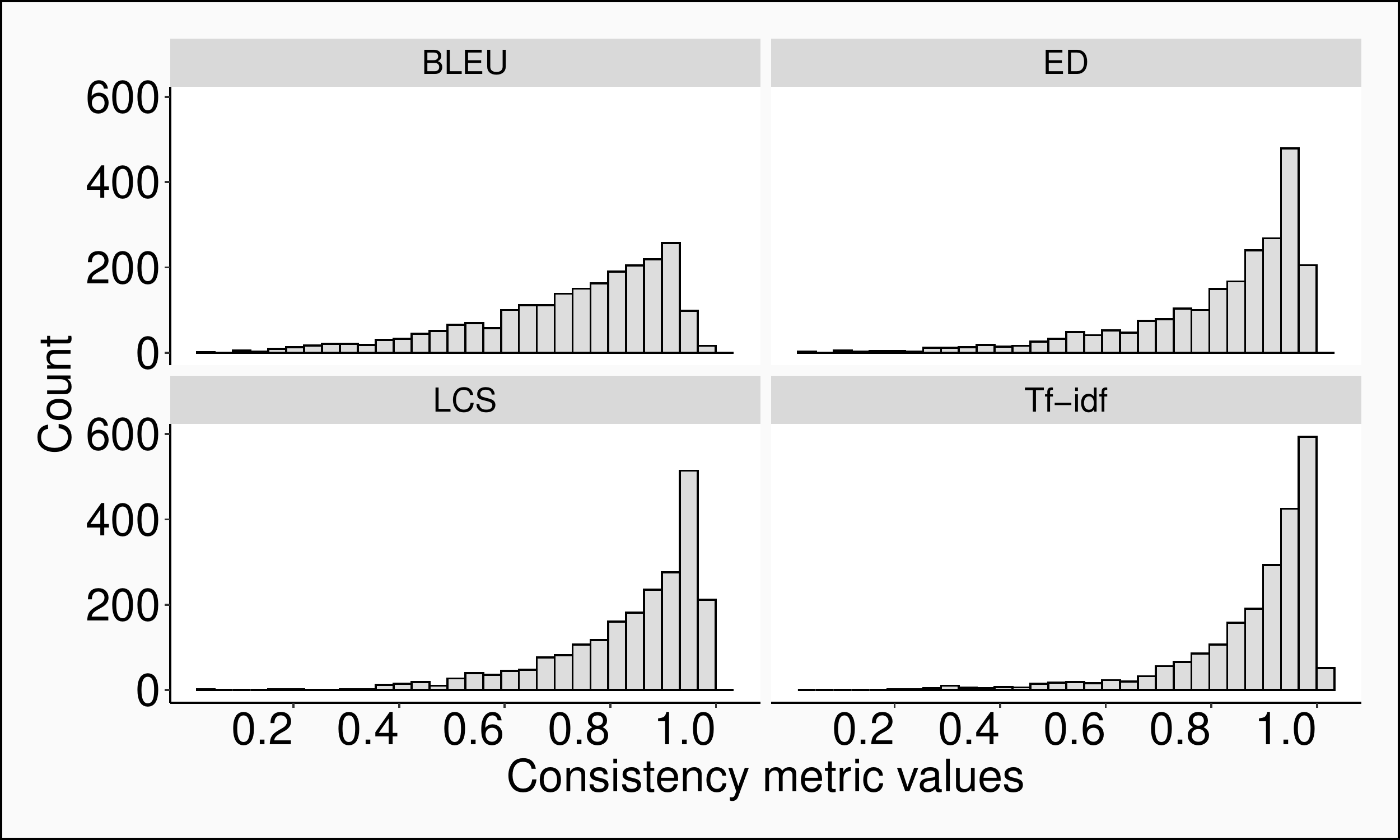}
    		\caption{Histogram of metric scores. A large number of mutant translations have similarity scores lower than one, indicating many inconsistent translations (RQ2).}
    		\label{fig:incondistribution}
\end{figure}

Figure~\ref{fig:incondistribution} shows the histogram of consistency scores that are lower than 1.0. 
As can be seen from the figure,
different metric values have different distributions, yet overall, all the four metrics report a large number of translations (i.e., around 47\% of the total translations) with 
a score below 1.0, indicating the presence of translation inconsistency.

Table~\ref{table:repair-numbug} shows the results of the reported inconsistent translations with different metric thresholds.
From Table~\ref{table:repair-numbug}, we can see that bugs remain even for highly permissive consistency thresholds.

\begin{table}[h!]\small
\caption{Number of reported inconsistency bugs with different thresholds between 1.0 and 0.6. 
With a 1.0 threshold the translation is deemed buggy if there is any detected inconsistency.
The lower the threshold, the more permissive is the criteria for deeming buggy.
As can be seen, bugs remain even for highly permissive consistency thresholds (RQ2).
}
\label{table:repair-numbug}
\vspace{-3mm}
\resizebox{.47\textwidth}{!}{
\begin{tabular}{@{}l|lrrrrr@{}}
\toprule
 & Thresh. & 1.0 & 0.9 & 0.8 & 0.7 & 0.6 \\ \midrule
 \multirow{4}{*}{\rotatebox{90}{GT}}
 & LCS & 2,053 (44\%)  & 865 (18\%)  & 342 (7\%)  & 123 (3\%)  & 57 (1\%) \\
 & ED & 2,053 (44\%)  & 913 (19\%)  & 401 (9\%)  & 198 (4\%)  & 101 (2\%) \\
 & Tf-idf & 2,459 (52\%)  & 548 (12\%)  & 208 (4\%)  & 71 (2\%)  & 21 (0\%) \\
 & BLEU & 2,053 (44\%)  & 1,621 (35\%)  & 911 (19\%)  & 510 (11\%)  & 253 (5\%) \\

 \midrule
\multirow{4}{*}{\rotatebox{90}{Transformer}} 
 & LCS & 2,213 (47\%)  & 1,210 (26\%)  & 634 (14\%)  & 344 (7\%)  & 184 (4\%) \\
 & ED & 2,213 (47\%)  & 1,262 (27\%)  & 700 (15\%)  & 428 (9\%)  & 267 (6\%) \\
 & Tf-idf & 2,549 (54\%)  & 851 (18\%)  & 399 (9\%)  & 188 (4\%)  & 112 (2\%) \\
 & BLEU & 2,213 (47\%)  & 1,857 (40\%)  & 1,258 (27\%)  & 788 (17\%)  & 483 (10\%) \\

 \bottomrule

\end{tabular}}
\end{table}

\paragraph{Manual Inspected Inconsistency}
In addition, we randomly sample 300 translations of the mutants.
Two of them do not parse so we use the remaining 298 translations for analysis.
For each mutant, the first two authors manually inspected its translation and the translation of the original sentence.
An inconsistency is reported when any of the following criteria are met: Apart from the mutated substitute word, the two translations 1) have different meanings; 2) have different tones; 3) use different characters for proper nouns.

Manual inspection reveals 107 (36\%) inconsistent translations for Google Translate, and 140 (47\%) inconsistent translations for Transformer\footnote{The Cohen's Kappa is 0.96/0.95 for Google Translate/Transformer, indicating that the inspection results are highly consistent.}.

\paragraph{Correlation between Metrics and Manual Inspection}
\label{sec:metrichumancorr}
We compare metric scores and human consistency assessment results. 
We split the 298 human-labelled translations into two groups based on manual inspection. 
One is labelled as consistent translations, the other is labelled as inconsistent translations.
We then check the metric value scores in each group. 

\begin{figure}[h]
    	    \centering
      		\includegraphics[width=0.95\linewidth]{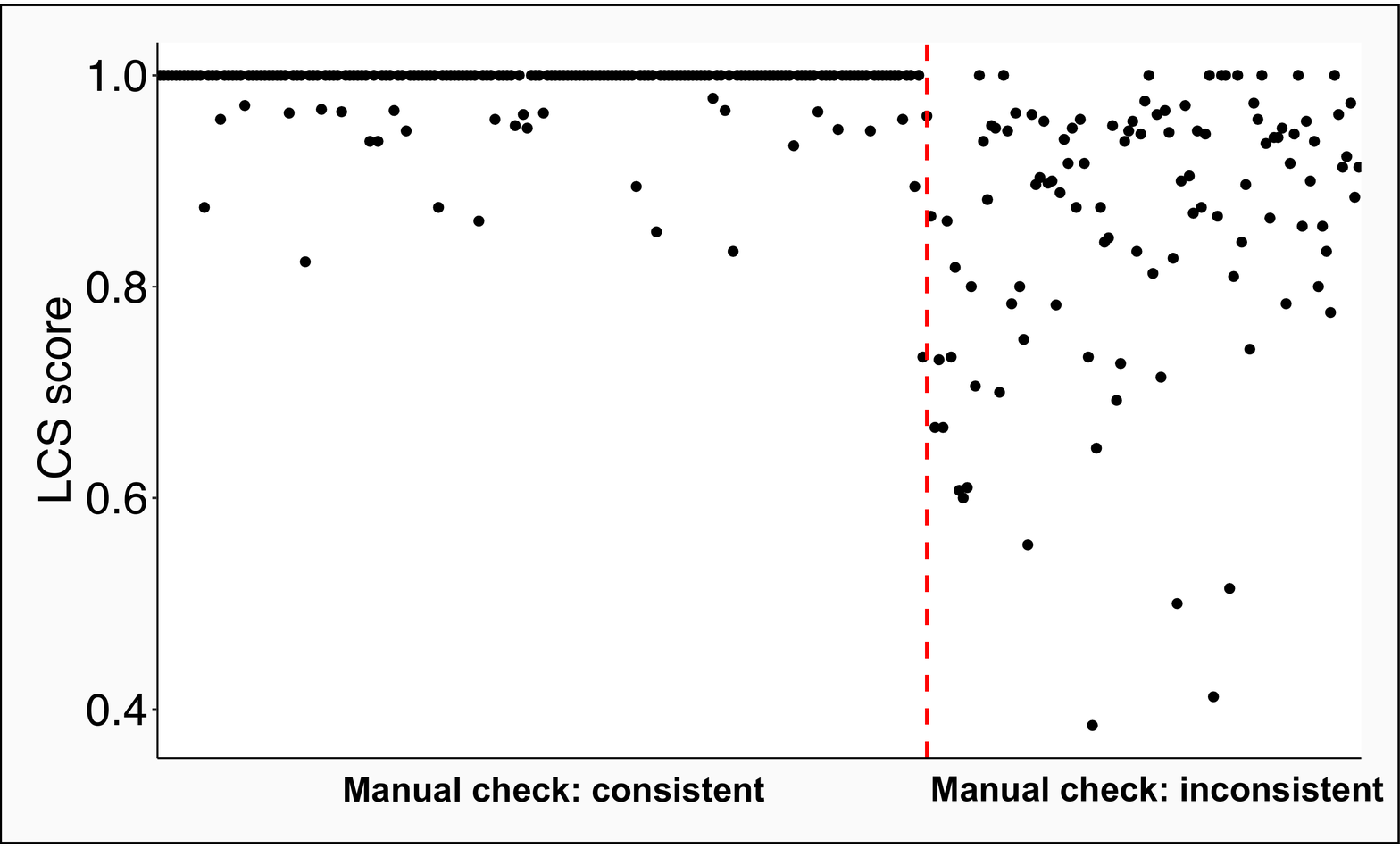}
      		\vspace{-5mm}
    		\caption{Comparison of metric scores and manual inspection of translation consistency. There is a good agreement between metric score values and human assessment (RQ2).}
    		\label{fig:human-metric-compare}
\end{figure}

Figure~\ref{fig:human-metric-compare} shows the results\footnote{This figure shows only the LCS metric. Full results are on our homepage~\cite{homepage}.}.
The points on the left/right of the dotted vertical line depict the metric values for the translations manually labelled as consistent/inconsistent.
We observe that most points (82.2\% ) in the left section have a score of 1.0.
The left section generally has higher score values (with a mean value of 0.99) than the right part (with a mean value of 0.86).
These observations indicate that the metric values and manual inspection tend to agree on translation inconsistency.
It is worth noting that Figure~\ref{fig:human-metric-compare} also shows some low metric values on the left section and high metric values on the right section, which indicates the presence of false positives and false negatives of using metrics to assess consistency.
We analyse false positives and false negatives in more detail below:



\paragraph{Threshold Learning}
Metric scores are continuous values. 
To automatically report inconsistency, 
we need to set a threshold for each metric.
In this paper, we choose a threshold that lets metric-value judgement best resemble manual inspection results.
To do this, we randomly sampled another 100 translations from Google Translate and manually label them as consistent or not.
We then used these 100 labels to choose a threshold (from 0.8 to 1.0 with a step of 0.01) with the largest F-measure score for each similarity metric.
The best thresholds for LCS, ED, tf-idf, and BLEU identified in this way are 0.963, 0.963, 0.999, 0.906 with F-measure scores 0.81, 0.82, 0.79, 0.82, respectively.
When a metric value falls below the so-identified threshold, our approach will report an inconsistency bug.

To know how well the chosen thresholds capture the boundary between consistency and inconsistency, we test the thresholds using the 298 previously sampled translations, on Google Translate and Transformer respectively.
The results are shown in Table~\ref{table:bug-precisionrecall}.
A false positive (FP in the table) means the threshold judges a translation as inconsistent but manual inspection is consistent. 
A false negative (FN in the table) means the threshold judges a translation as consistent but manual inspection is inconsistent. 
From the table, the proportion of false positives and false negatives are all below 10\%, which we deem to be acceptable.

After a manual check of FPs and FNs, we found that an FN may happen when there is a minor character difference, but with a different meaning or tone. For example, in our results, one mutant translation has an extra \emph{er} (means ``But'') which does not exist in the original translation. Manual inspection deemed this to be inconsistent, while metrics did not.
An FP may happen when there are many different words in the two translations, but they share the same meaning. For example, Chinese phrases \emph{shang wei} and \emph{hai mei you} both mean ``not yet'', but the characters that express each phrase are totally different.

The harm that an FP in the testing process may bring lies in the possibility that our approach may make the translation worse. Section~\ref{sec:improveassessedbyhuman} explores this possibility.


\begin{table}[h!]\small
 \caption{Precision/recall for inconsistency revealing (RQ2)}
    		\label{table:bug-precisionrecall}
    		\vspace{-3mm}
 \resizebox{.48\textwidth}{!}{
\begin{tabular}{@{}l|lrrrrrrrrr@{}}
\toprule
& Metric & TN& FN & FP&TP&Precision&Recall& F-meas.\\ \midrule
\multirow{4}{*}{\rotatebox{90}{GT}}
&LCS & 169 (57\%)  & 16 (5\%)  & 22 (7\%)  & 91 (31\%)  & 0.81 & 0.85 & 0.83\\
&ED & 169 (57\%)  & 16 (5\%)  & 22 (7\%)  & 91 (31\%)  & 0.81 & 0.85 & 0.83\\
&tf-idf & 162 (54\%)  & 12 (4\%)  & 29 (10\%)  & 95 (32\%)  & 0.77 & 0.89 & 0.82\\
&BLEU & 171 (57\%)  & 20 (7\%)  & 20 (7\%)  & 87 (29\%)  & 0.81 & 0.81 & 0.81\\

\midrule
\multirow{4}{*}{\rotatebox{90}{Transformer}}
&LCS & 142 (48\%)  & 22 (7\%)  & 16 (5\%)  & 118 (40\%)  & 0.88 & 0.84 & 0.86\\
&ED & 142 (48\%)  & 21 (7\%)  & 16 (5\%)  & 119 (40\%)  & 0.88 & 0.85 & 0.87\\
&tf-idf & 141 (47\%)  & 11 (4\%)  & 17 (5\%)  & 129 (43\%)  & 0.88 & 0.92 & 0.90\\
&BLEU & 147 (49\%)  & 23 (8\%)  & 11 (4\%)  & 117 (39\%)  & 0.91 & 0.84 & 0.87\\
 \bottomrule
\end{tabular}
}
\end{table}

\paragraph{Overall Number of Inconsistency Issues}
After identifying the thresholds, we apply them to the translations of the 4,592 generated inputs\footnote{We removed those 100 translations used for threshold learning from our test set and used the remaining 4,592 inputs to conduct testing.}, and check how many of them fall below the threshold.
It turns out that for Transformer, the inconsistent translation results are LCS: 1,917 (42\%); ED: 1,923 (42\%); tf-idf: 2,102 (46\%); BLEU: 1,857 (40\%). 
Thus, overall, about two fifths of the translations fall below our chosen consistency thresholds.
For Google Translate, the inconsistent translation results are LCS: 1,708 (37\%); ED: 1,716 (37\%); tf-idf: 1,875 (41\%); BLEU: 1,644 (36\%). 
This also shows that Google Translate is slightly better than Transformer with respect to consistency.

\begin{tcolorbox}
Answers to \textbf{RQ2}: 
The metrics have an F-measure of 0.82/0.88 when detecting inconsistency bugs for Google Translate/Transformer.
Both metrics and manual inspection reveal that Google Translate has approximately 36\% inconsistent translations on \techname test inputs. \end{tcolorbox} 

\subsection{Bug-repair Ability (RQ3)}

\subsubsection{Improvement Assessed by Metrics}
We apply our repair approaches to all the inconsistent translations, and check how many translations can be repaired with our approach. 
For each sentence, we generate 16 mutants for repair (we study the influence of the number of mutants for repair in Section~\ref{sec:mutantnumberinfluence}).

Table~\ref{table:repair-bugnumreduce} shows the results, 
where each cell contains the number and proportion of inconsistency bugs that the repair approach repairs. 
The Column ``Probability'' represents the results for probability-reference (grey-box repair); 
the Columns ``Cross-reference'' represents the results for cross-reference (black-box repair); 
For Google translate, since the grey-box approach is not applicable, we only present the results for the black-box approach.

From the table, \techname reduces, on average, 28\% of bugs for the black-box approach in Google Translate. 
For the Transformer model, we can see the grey-box approach repairs 30\% of bugs,  
the black-box one repairs 19\% to 20\% of bugs.
These experimental results show that the grey-box and black-box approaches are effective at repairing inconsistency bugs. 

\begin{table}[]\small
\caption{Number and proportion of repaired bugs (RQ3).}
\label{table:repair-bugnumreduce}
\vspace{-3mm}
 \resizebox{.39\textwidth}{!}{
\begin{tabular}{@{}l|p{1.4cm}p{2.5cm}r@{}}
\toprule
& Metric & Probability & Cross-reference \\ \midrule
\multirow{4}{*}{{GT}}
& LCS & -- & 493 (28\%) \\
 & ED & --  & 503 (29\%) \\
 & tf-idf & --  & 478 (25\%) \\
 & BLEU & -- & 484 (29\%)\\
\midrule
\multirow{4}{*}{{Transformer}}
 & LCS & 583 (30\%) & 374 (19\%) \\
 & ED & 581 (30\%) & 389 (20\%)  \\
 & tf-idf & 640 (30\%) & 400 (19\%) \\
 & BLEU & 565 (30\%) & 380 (20\%)\\

 \bottomrule
\end{tabular}
}

\end{table}



    	
\subsubsection{Improvement Assessed by Human}
\label{sec:improveassessedbyhuman}

Program repair studies typically include a manual assessment process~\cite{Xin:2017:LSC:3155562.3155644,Saha:2017:EEO:3155562.3155643,Jiang:2018:SPR:3213846.3213871} in order to validate their findings. Following a similar practice, 
the first two authors (manually) checked the repaired results of the previously labelled 298 sampled translations.
The goal was to check whether the changes in translations patched by our repair approach improve translation consistency. 
Since our repair may also improve translation acceptability, we also check whether the changes bring any translation acceptability improvement.
For cross-reference based repair, we manually assessed only the LCS metric since our (previous) results showed similar results among the other metrics.

Among the 298 sentence pairs, 113/136 of them are reported as having translation inconsistency on Google Translate/Transformer by metrics. 
Our repair thus targets all these sentence pairs, including the translations of both the original sentence and the mutant.
The probability-based repair approach finally changed
58 (out of 136) pairs for Transformer;
The cross-reference-based repair approach finally changed 39/27 (out of 113/136) pairs for Google Translate/Transformer.

For the translation pairs that have been modified by \techname, the first two authors then manually compared two dimensions: 
1) the translation consistency before and after repair;
2) the acceptability of the translations (for both the original sentence and the mutant) before and after repair.
For each dimension, the authors gave labels of ``Improved'', ``Unchanged'', or ``Decreased'', considering both adequacy and fluency (see explanations of these two terms in Section~\ref{sec:testoracleidentification})\footnote{The mean Kappa score is for labelling translation consistency between the two human labels,
0.97 for labelling the translation of the original sentence,
and 0.81 for labelling the translation of the mutant sentence.}.

Table~\ref{tab:repairperformance-humanassess} shows the results.
The first four rows are for Google Translate.
The remaining rows are for Transformer. 
The rows with ``overall'' show results of translation acceptability improvement among the translations for both original sentences and mutant sentences.
The rows with ``original''/``mutant'' show the translation repair results for original/mutant sentences.

We observe that \techname has a good effectiveness in improving translation consistency.
For example, on average, 87\% translation pairs improve the consistency for Google Translate and Transformer, while all together we observe only 3 translation consistency decreases.
We checked these decreased-consistency repairs and found that, for one case, the original sentence translation has been improved but not for the mutant translation, and thus after repair, the improved translation of the original sentence does not match well with the unimproved translation of the mutant. The remaining two cases arise because the repairs of the original sentences decreased, while our approach did not touch the mutant translations. 

The main purpose of our repair approach is to improve translation consistency.
Translation acceptability improvement is a ``bonus'' of our approach.
From Table~\ref{tab:repairperformance-humanassess}, perhaps to our surprise, the repair approach improves the translation acceptability for around one fourth (27\%) repairs.
There are 8\% repairs with decreased acceptability. 
Based on our manual inspection, the reason for decreased acceptability is that occasionally, the repair approach may trade quality for consistency.

\begin{tcolorbox}
Answers to \textbf{RQ3}: Black-box repair reduces on average 28\%/19\% bugs for Google Translate/Transformer.
Grey-box repair reduces on average 30\% bugs for Transformer.
Human inspection indicates that the repaired translations improve consistency in 87\% of the cases (reducing it in 2\%), and have better translation acceptability in 27\% of the cases (worse in 8\%).
\end{tcolorbox}

\begin{table}[]
    		\centering
\caption{Improvement Based on Manual Inspection (RQ3)}
\label{tab:repairperformance-humanassess}
\vspace{-3mm}
\resizebox{.47\textwidth}{!}{
\begin{tabular}{@{}l|lrrr@{}}
\toprule
& Aspect & Improved & Unchanged & Decreased \\ 
\midrule
\multirow{4}{*}{\rotatebox{90}{GT.LCS}}
&\textbf{Translation consistency} & 33 (85\%) & 4 (10\%) & 2 (5\%) \\
&\textbf{Translation acceptability: overall} & 22 (28\%) & 48 (62\%) & 8 (10\%) \\
&Translation acceptability: original & 10 (26\%) & 23 (59\%) & 6 (15\%) \\
&Translation acceptability: mutant & 12 (31\%) & 25 (64\%) & 2 (5\%) \\ \midrule
\multirow{4}{*}{\rotatebox{90}{Trans.LCS}}
&\textbf{Translation consistency} & 24 (89\%) & 3 (11\%) & 0 (0\%) \\
&\textbf{Translation acceptability: overall} & 15 (28\%) & 37 (69\%) & 2 (4\%) \\
&Translation acceptability: original & 7 (26\%) & 19 (70\%) & 1 (4\%) \\
&Translation acceptability: mutant & 8 (30\%) & 18 (67\%) & 1 (4\%) \\
\midrule
\multirow{4}{*}{\rotatebox{90}{Trans.Prob}}
&\textbf{Translation consistency} & 51 (88\%) & 6 (10\%) & 1 (2\%) \\
&\textbf{Translation acceptability: overall} & 30 (26\%) & 76 (66\%) & 10 (9\%) \\
&Translation acceptability: original & 15 (26\%) & 36 (62\%) & 7 (12\%) \\
&Translation acceptability: mutant & 15 (26\%) & 40 (69\%) & 3 (5\%) \\

\bottomrule
\end{tabular}}
\end{table}

\section{Extended Analysis and Discussion}
This section provides further details and analysis.

\paragraph{\textbf{Example of Repaired Translations}}
Table~\ref{table:showcase} gives some examples of our improvement of mistranslation.
The first column is the translation input; the second column shows the original translation output (converted to Pinyin), where words in italic explain the mistranslated parts; the last column shows our improved translation.

\begin{table*}[]
\caption{Examples of Repaired Translations.}
\vspace{-3mm}
\label{table:showcase}
\resizebox{.98\textwidth}{!}{
\begin{tabular}{@{}p{7cm}p{6.5cm}p{7cm}@{}}\toprule
\textbf{Input}  & \textbf{Original translation}  & \textbf{Repaired translation}\\ \midrule
Female students do \textbf{good} research in computer science.                      & \pinyin{nv5}xuesheng zai jisuanji kexue fangmian zuole \textbf{henduo} yanjiu \emph{\jie{[Bug: ``good'' $\rightarrow$ ``a lot''.]}}              & \pinyin{nv5}xueshengzai jisuanji kexue fangmian zuole \textbf{henhaode} yanjiu                                                       \\\midrule
If you need help, you can enjoy timely services by pressing a nearby \textbf{one of} the 41 call buttons in the station. 
& ruguo ni xuyao bangzhu, ni keyi tongguo an fujin de 41 ge hujiao anniu xiangshou jishi de fuwu. \emph{\jie{[Bug: ``one of'' is not translated.]}}
& ruguo ni xuyao bangzhu, ni keyi tongguo an fujin de 41 ge hujiao anniu \textbf{zhong de yige} lai xiangshou jishi de fuwu. \\\midrule
Original Title : Canada Police Kill \textbf{IS} Supporters : Preparations for Homemade Bomb Downtown Attack Near the End.  
& yuanshi biaoti: jianada jingcha sha le \textbf{wo de} zhichizhe: wei hema zhizao baozha de zhunbei. \emph{\jie{[Bug: ``IS'' $\rightarrow$ ``my''.]}}
& yuanshi biaoti: jianada jingcha shalu \textbf{IS} zhichizhe: wei hema zhizao zhadan de zhunbei.
\\\midrule
Ban political campaigners and activists from handling \textbf{completed} postal votes and postal vote envelopes. 
& jinzhi zhengzhi jingsuanzhe he huodongfenzi chuli \textbf{wan} de youzheng xuanpiao he youzheng xuanpiao xifeng \emph{\jie{[Bug: ``completed'' is mistranslated.]}}
& jinzhi zhengzhi jingsuanzhe he huodongfenzi chuli \textbf{yi wancheng} de youzheng xuanpiao he youzheng xuanpiao xifeng.
\\  
\bottomrule
\end{tabular}}
\end{table*}

\paragraph{\textbf{Effectiveness and Efficiency of Data Augmentation}}

Previous work has adopted data augmentation to improve the robustness of machine learning models~\cite{cheng2018towards,ribeiro2018semantically}.
In our work, concerning translators whose source code is known, training data augmentation is also a candidate solution to increase translation consistency.

\begin{figure}[h]\small
    \center
\includegraphics[scale=0.31] {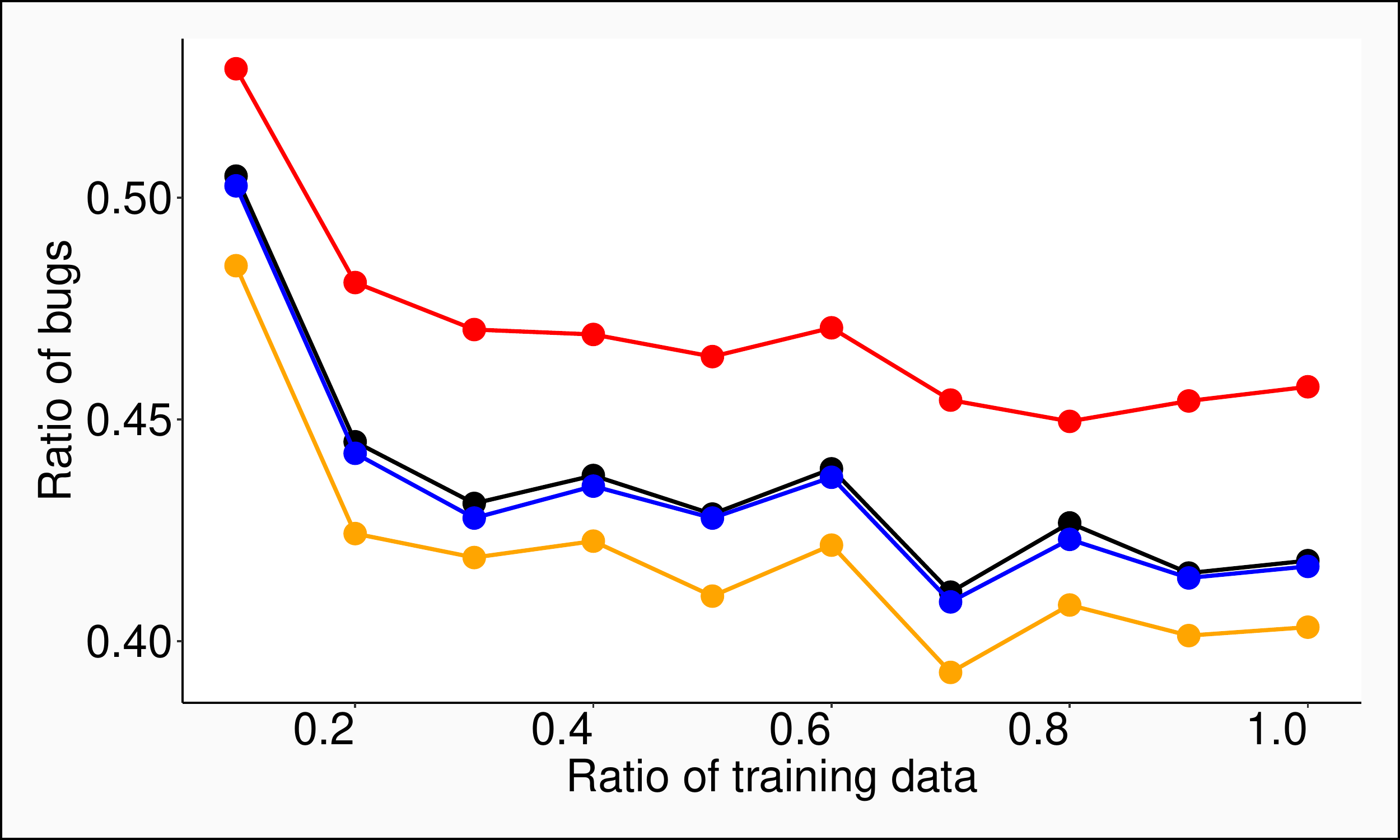}
                        \vspace{-2mm}
    \caption{Ratio of inconsistency bugs with different training-data size for Transformer.}
    \label{fig:difftrainingsize}
\end{figure}

To investigate this option, we designed experiments to study whether adding more training data would yield better translation consistency. 
We controlled the training data size and used 10\%, 20\%, ..., and 90\% of the original training data to build Transformer respectively. 
Figure~\ref{fig:difftrainingsize} shows the results.
When the size of the training data ratio is between 0.7 and 1.0, we did not observe a decreasing trend.
This indicates that augmenting training data may have limited effectiveness in improving translation inconsistency.

Data augmentation needs model retraining.
We found that using 100\% training data to train the translator model took as much as 19 hours under our current experimental configuration (see more details in Section~\ref{sec:experimentalsetup}).
In practice, data collection, labelling, and processing also needs time.
All together, we find little evidence that augmentation is a complete solution to this translation repair problem.

Compared with model retraining approaches like training data augmentation, \techname has the following advantages:
\textbf{1)} \techname require neither the source code nor training data, and is either completely black-box or requires only predictive probability (grey-box);
\textbf{2)} \techname can have lower repair cost since it does not need additional data and does not require model retraining;
\textbf{3)} \techname is more flexible, because it enables the repair of a specific bug without touching other well-formed translations.

\paragraph{\textbf{Efficiency of \techname}}

The cost of \techname includes both testing and repair. 
Under our current experimental configuration (see more details in Section~\ref{sec:experimentalsetup}), the mean time for testing is 0.97s per sentence;
the mean time for repair is 2.68s per sentence for the probability-based approach, and 2.93s per sentence for the cross-reference-based approach.
Thus, with current experimental configuration, when using \techname to optimise the real-time online machine translation,
for a sentence that is deemed non-buggy, it would take less than 1 second for the end user to receive a final translation. 
For a sentence that is deemed buggy, it would take less than 4 seconds (testing and repair) to get a final translation.

\paragraph{\textbf{Influence of Mutant Number}}
\label{sec:mutantnumberinfluence}

We generate mutants during both inconsistency testing and repair.
For the test/repair process, our default configuration generates at most 5/16 mutants for each sentence. 
To investigate how the number of mutants affects the testing and repair performance,
we repeat our experiments with 1 or 3 mutants for test generation, and with 4 or 8 mutants for repair. 
We then compare the number of revealed inconsistency bugs and the number of bugs that our approaches repair. 
For repair, we only present results of grey-box repair approach in the paper due to space reason, as shown by Table~\ref{table:mutantinfluence}.
The full results are available on our homepage~\cite{homepage}.
We observe that during testing, using more mutants helps to reveal more inconsistency bugs.
It is the same for repair, but using 4 mutants during repair also has an acceptable effectiveness.

\begin{table}[]
\caption{Number of detected and repaired bugs with different number of mutants. }
\label{table:mutantinfluence}
\vspace{-3mm}
\resizebox{.45\textwidth}{!}{
\begin{tabular}{p{1.5cm}|ccc|ccc}\toprule
\multirow{2}{*}{{Metric}}&\multicolumn{3}{c|}{Mutant number for testing}&\multicolumn{3}{|c}{Mutant number for repair}\\ \cmidrule{2-7}
&1&3&5&4&8&16\\ \midrule
LCS & 535 & 1,345 & 1,917 & 490 & 551 & 583\\
ED & 536 & 1,349 & 1,923 & 488 & 549 & 581\\
tf-idf &  583 & 1,467 & 2,102 & 534 & 600 & 640\\
BLEU & 513 & 1,300 & 1,857 & 483 & 538 & 565\\
\bottomrule
\end{tabular}
}
\end{table}

\paragraph{\textbf{Application Scenario}}

\techname can be applied end to end.
Given a translation input and a machine translator, our approach will automatically test and repair the translation output, and give a new translation output to the end user.

\section{Related Work}

Software testing research has typically targeted traditional (non-machine-learning-based) software systems. 
However, the recent rise in the real-world importance of machine learning has resulted in a concomitant rise in the level of research activity in software testing for machine learning~\cite{zhang2019machine}.
At the same time, software repair concepts and techniques remain relatively under explored for machine learning systems.
In this section, we review the relationship of our proposed machine translation testing and repair with previous work on testing and repair machine translation systems, which mainly focus on translation robustness. 

\paragraph{\textbf{Translation Robustness Testing}}
To test translation robustness, researchers have explored how perturbations on the test inputs affect translations.
Heigold et al.~\cite{DBLP:conf/amta/HeigoldVNG18} studied three types of character-level noisy test inputs that are generated via character swapping, word scrambling, and character flipping.
They found that machine translation systems are very sensitive to slightly perturbed sentences that do not pose a challenge to humans.
Belinkov and Bisk~\cite{belinkov2017synthetic} had a similar conclusion, not only on synthetic noise, but also on natural noise (naturally occurring errors like typos and misspellings). 
To have more diverse test cases for robustness testing, Zhao et al.~\cite{DBLP:journals/corr/abs-1710-11342} used Generative Adversarial Networks (GANs)~\cite{NIPS2014_5423}. 
They projected the input sentence to a latent space, which is then used for searching for test sentences close to the input. 

These work targets robustness testing. 
The test inputs are synthetic errors or naturally-occurring errors.
The test oracles are usually BLEU scores, which are bounded with human oracles. 
Our approach targets translation consistency, and we generate consistent test inputs by context-similar word replacement, instead of involving errors.
Our approach also does not require human oracles during testing.

Sun and Zhou~\cite{sun2018metamorphic} proposed metamorphic relations for machine translation.
There are two major differences between our work and theirs:
1) their work concerns testing only; we also repair;
2) their test input generation merely replaces human names before ``likes'' or ``hates'' and brands after them; our approach is comparatively more exhaustive.

\paragraph{\textbf{Translation Robustness Improvement}}

To improve translation robustness, previous work relies largely on data augmentation, i.e., to add noisy data into the training data and to retrain the model. 
Some work used model-independent data generation (also called black-box data generation).
Heigold et al.~\cite{DBLP:conf/amta/HeigoldVNG18},
Belinkov and Bisk~\cite{belinkov2017synthetic}, and Sperber et al.~\cite{sperber2017toward} used synthetic noise to retain the model.
Karpukhin et al.~\cite{karpukhin2019training} evaluated the impact of percentages of synthetic noise to the training set. 

Some work uses model-dependent data generation (white-box data generation).
Ebrahimi et al.~\cite{ebrahimi-etal-2018-hotflip} introduced an approach that relies on an atomic flip operation. This operation generates tests by swapping characters based on the gradients of the input vectors. 
Cheng et al.~\cite{DBLPconfaclChengJM19} proposed a gradient-based method to generate adversarial sentences by conducting word replacement. 

There is also work on improving robustness via optimising the learning algorithms.
Belinkov and Bisk~\cite{belinkov2017synthetic} proposed to use a structure-invariant representation for synthetic noise in the network. They find that a character CNN representation is more robust than others. 
Cheng et al.~\cite{cheng2018towards} introduced stability training by adding a new component for discriminating the noise in the training set. 
This component reduces the impact of the noise, and yields more stable translations when making synonymous perturbations.

These previous approaches target overall robustness improvement for all translations, rather than fixing specific mistranslations.

\section{Conclusion}

In this paper, we presented \techname, the first approach that automatically tests and improves context-similar translation consistency.
\techname takes a sentence and applies a context-similar mutation to generate slightly altered (mutated) sentences, to be used to test machine translation systems.
Testing is performed by translating and comparing the original with the mutated sentences. 
To judge consistency, \techname calculates the similarity of the translation subsequences. 
When context-similar mutations yield above-threshold disruption to the translation of the unchanged part, \techname deems this to be a potential bug. 
In addition to testing, \techname also automatically repairs inconsistencies in a black-box or grey-box manner, which post-processes the translations with reference to the translations of mutated sentences.

\newpage
\balance
\bibliography{ase}


\begin{thebibliography}{53}


\ifx \showCODEN    \undefined \def \showCODEN     #1{\unskip}     \fi
\ifx \showDOI      \undefined \def \showDOI       #1{#1}\fi
\ifx \showISBNx    \undefined \def \showISBNx     #1{\unskip}     \fi
\ifx \showISBNxiii \undefined \def \showISBNxiii  #1{\unskip}     \fi
\ifx \showISSN     \undefined \def \showISSN      #1{\unskip}     \fi
\ifx \showLCCN     \undefined \def \showLCCN      #1{\unskip}     \fi
\ifx \shownote     \undefined \def \shownote      #1{#1}          \fi
\ifx \showarticletitle \undefined \def \showarticletitle #1{#1}   \fi
\ifx \showURL      \undefined \def \showURL       {\relax}        \fi
\providecommand\bibfield[2]{#2}
\providecommand\bibinfo[2]{#2}
\providecommand\natexlab[1]{#1}
\providecommand\showeprint[2][]{arXiv:#2}

\bibitem[\protect\citeauthoryear{{ Ralph Weischedel, Martha Palmer, Mitchell
  Marcus, Eduard Hovy, Sameer Pradhan, Lance Ramshaw, Nianwen Xue, Ann Taylor,
  Jeff Kaufman, Michelle Franchini, Mohammed El-Bachouti, Robert Belvin, Ann
  Houston}}{{ Ralph Weischedel, Martha Palmer, Mitchell Marcus, Eduard Hovy,
  Sameer Pradhan, Lance Ramshaw, Nianwen Xue, Ann Taylor, Jeff Kaufman,
  Michelle Franchini, Mohammed El-Bachouti, Robert Belvin, Ann
  Houston}}{2013}]%
        {OntoNotes}
\bibfield{author}{\bibinfo{person}{{ Ralph Weischedel, Martha Palmer, Mitchell
  Marcus, Eduard Hovy, Sameer Pradhan, Lance Ramshaw, Nianwen Xue, Ann Taylor,
  Jeff Kaufman, Michelle Franchini, Mohammed El-Bachouti, Robert Belvin, Ann
  Houston}}.} \bibinfo{year}{2013}\natexlab{}.
\newblock \bibinfo{title}{{OntoNotes}}.
\newblock
  \bibinfo{howpublished}{\url{https://catalog.ldc.upenn.edu/LDC2013T19}}.
\newblock


\bibitem[\protect\citeauthoryear{Anonymous}{Anonymous}{2019}]%
        {homepage}
\bibfield{author}{\bibinfo{person}{Anonymous}.}
  \bibinfo{year}{2019}\natexlab{}.
\newblock \bibinfo{title}{{TransRepair Homepage}}.
\newblock
  \bibinfo{howpublished}{\url{https://github.com/anonymous54351/TransRepair}}.
\newblock


\bibitem[\protect\citeauthoryear{Belinkov and Bisk}{Belinkov and Bisk}{2018}]%
        {belinkov2017synthetic}
\bibfield{author}{\bibinfo{person}{Yonatan Belinkov} {and}
  \bibinfo{person}{Yonatan Bisk}.} \bibinfo{year}{2018}\natexlab{}.
\newblock \showarticletitle{Synthetic and natural noise both break neural
  machine translation}. In \bibinfo{booktitle}{\emph{Proc. ICLR}}.
\newblock


\bibitem[\protect\citeauthoryear{Chen, Cheung, and Yiu}{Chen
  et~al\mbox{.}}{1998}]%
        {chen1998metamorphic}
\bibfield{author}{\bibinfo{person}{Tsong~Y Chen}, \bibinfo{person}{Shing~C
  Cheung}, {and} \bibinfo{person}{Shiu~Ming Yiu}.}
  \bibinfo{year}{1998}\natexlab{}.
\newblock \bibinfo{booktitle}{\emph{Metamorphic testing: a new approach for
  generating next test cases}}.
\newblock \bibinfo{type}{{T}echnical {R}eport}.
\newblock


\bibitem[\protect\citeauthoryear{Cheng, Jiang, and Macherey}{Cheng
  et~al\mbox{.}}{2019}]%
        {DBLPconfaclChengJM19}
\bibfield{author}{\bibinfo{person}{Yong Cheng}, \bibinfo{person}{Lu Jiang},
  {and} \bibinfo{person}{Wolfgang Macherey}.} \bibinfo{year}{2019}\natexlab{}.
\newblock \showarticletitle{Robust Neural Machine Translation with Doubly
  Adversarial Inputs}. In \bibinfo{booktitle}{\emph{Proceedings of the 57th
  Conference of the Association for Computational Linguistics, {ACL} 2019,
  Florence, Italy, July 28- August 2, 2019, Volume 1: Long Papers}}.
  \bibinfo{pages}{4324--4333}.
\newblock
\urldef\tempurl%
\url{https://www.aclweb.org/anthology/P19-1425/}
\showURL{%
\tempurl}


\bibitem[\protect\citeauthoryear{Cheng, Tu, Meng, Zhai, and Liu}{Cheng
  et~al\mbox{.}}{2018}]%
        {cheng2018towards}
\bibfield{author}{\bibinfo{person}{Yong Cheng}, \bibinfo{person}{Zhaopeng Tu},
  \bibinfo{person}{Fandong Meng}, \bibinfo{person}{Junjie Zhai}, {and}
  \bibinfo{person}{Yang Liu}.} \bibinfo{year}{2018}\natexlab{}.
\newblock \showarticletitle{Towards robust neural machine translation}.
\newblock \bibinfo{journal}{\emph{arXiv preprint arXiv:1805.06130}}
  (\bibinfo{year}{2018}).
\newblock


\bibitem[\protect\citeauthoryear{{CWMT}}{{CWMT}}{2018}]%
        {CWMT}
\bibfield{author}{\bibinfo{person}{{CWMT}}.} \bibinfo{year}{2018}\natexlab{}.
\newblock \bibinfo{title}{{The CWMT Dataset}}.
\newblock \bibinfo{howpublished}{\url{http://nlp.nju.edu.cn/cwmt-wmt/}}.
\newblock


\bibitem[\protect\citeauthoryear{Doddington}{Doddington}{2002}]%
        {doddington2002automatic}
\bibfield{author}{\bibinfo{person}{George Doddington}.}
  \bibinfo{year}{2002}\natexlab{}.
\newblock \showarticletitle{Automatic evaluation of machine translation quality
  using n-gram co-occurrence statistics}. In
  \bibinfo{booktitle}{\emph{Proceedings of the second international conference
  on Human Language Technology Research}}. Morgan Kaufmann Publishers Inc.,
  \bibinfo{pages}{138--145}.
\newblock


\bibitem[\protect\citeauthoryear{Eberhard, Simons, and Fennig}{Eberhard
  et~al\mbox{.}}{2019}]%
        {eberhard2019ethnologue}
\bibfield{author}{\bibinfo{person}{David~M Eberhard}, \bibinfo{person}{Gary~F
  Simons}, {and} \bibinfo{person}{Charles~D Fennig}.}
  \bibinfo{year}{2019}\natexlab{}.
\newblock \showarticletitle{Ethnologue: Languages of the world}.
\newblock  (\bibinfo{year}{2019}).
\newblock


\bibitem[\protect\citeauthoryear{Ebrahimi, Rao, Lowd, and Dou}{Ebrahimi
  et~al\mbox{.}}{2018}]%
        {ebrahimi-etal-2018-hotflip}
\bibfield{author}{\bibinfo{person}{Javid Ebrahimi}, \bibinfo{person}{Anyi Rao},
  \bibinfo{person}{Daniel Lowd}, {and} \bibinfo{person}{Dejing Dou}.}
  \bibinfo{year}{2018}\natexlab{}.
\newblock \showarticletitle{{H}ot{F}lip: White-Box Adversarial Examples for
  Text Classification}. In \bibinfo{booktitle}{\emph{Proceedings of the 56th
  Annual Meeting of the Association for Computational Linguistics (Volume 2:
  Short Papers)}}. \bibinfo{publisher}{Association for Computational
  Linguistics}, \bibinfo{address}{Melbourne, Australia},
  \bibinfo{pages}{31--36}.
\newblock
\urldef\tempurl%
\url{https://doi.org/10.18653/v1/P18-2006}
\showDOI{\tempurl}


\bibitem[\protect\citeauthoryear{Foundation}{Foundation}{2019}]%
        {wdiff}
\bibfield{author}{\bibinfo{person}{Free~Software Foundation}.}
  \bibinfo{year}{2019}\natexlab{}.
\newblock \bibinfo{title}{GNU Wdiff}.
\newblock
\newblock
\urldef\tempurl%
\url{https://www.gnu.org/software/wdiff/}
\showURL{%
\tempurl}


\bibitem[\protect\citeauthoryear{Giglio and Caulk}{Giglio and Caulk}{1965}]%
        {giglio:uccialli}
\bibfield{author}{\bibinfo{person}{Carlo Giglio} {and} \bibinfo{person}{Richard
  Caulk}.} \bibinfo{year}{1965}\natexlab{}.
\newblock \showarticletitle{Article 17 of the Treaty of Uccialli}.
\newblock \bibinfo{journal}{\emph{The Journal of African History}}
  \bibinfo{volume}{6}, \bibinfo{number}{2} (\bibinfo{year}{1965}),
  \bibinfo{pages}{221--231}.
\newblock


\bibitem[\protect\citeauthoryear{Goldberg and Levy}{Goldberg and Levy}{2014}]%
        {goldberg2014word2vec}
\bibfield{author}{\bibinfo{person}{Yoav Goldberg} {and} \bibinfo{person}{Omer
  Levy}.} \bibinfo{year}{2014}\natexlab{}.
\newblock \showarticletitle{word2vec Explained: deriving Mikolov et al.'s
  negative-sampling word-embedding method}.
\newblock \bibinfo{journal}{\emph{arXiv preprint arXiv:1402.3722}}
  (\bibinfo{year}{2014}).
\newblock


\bibitem[\protect\citeauthoryear{Goodfellow, Pouget-Abadie, Mirza, Xu,
  Warde-Farley, Ozair, Courville, and Bengio}{Goodfellow et~al\mbox{.}}{2014}]%
        {NIPS2014_5423}
\bibfield{author}{\bibinfo{person}{Ian Goodfellow}, \bibinfo{person}{Jean
  Pouget-Abadie}, \bibinfo{person}{Mehdi Mirza}, \bibinfo{person}{Bing Xu},
  \bibinfo{person}{David Warde-Farley}, \bibinfo{person}{Sherjil Ozair},
  \bibinfo{person}{Aaron Courville}, {and} \bibinfo{person}{Yoshua Bengio}.}
  \bibinfo{year}{2014}\natexlab{}.
\newblock \showarticletitle{Generative Adversarial Nets}.
\newblock In \bibinfo{booktitle}{\emph{Advances in Neural Information
  Processing Systems 27}}, \bibfield{editor}{\bibinfo{person}{Z.~Ghahramani},
  \bibinfo{person}{M.~Welling}, \bibinfo{person}{C.~Cortes},
  \bibinfo{person}{N.~D. Lawrence}, {and} \bibinfo{person}{K.~Q. Weinberger}}
  (Eds.). \bibinfo{publisher}{Curran Associates, Inc.},
  \bibinfo{pages}{2672--2680}.
\newblock
\urldef\tempurl%
\url{http://papers.nips.cc/paper/5423-generative-adversarial-nets.pdf}
\showURL{%
\tempurl}


\bibitem[\protect\citeauthoryear{{Google}}{{Google}}{2019}]%
        {googletranslate}
\bibfield{author}{\bibinfo{person}{{Google}}.} \bibinfo{year}{2019}\natexlab{}.
\newblock \bibinfo{title}{Google Translate}.
\newblock \bibinfo{howpublished}{\url{http://translate.google.com}}.
\newblock


\bibitem[\protect\citeauthoryear{Graham, Baldwin, Harwood, Moffat, and
  Zobel}{Graham et~al\mbox{.}}{2012}]%
        {graham2012measurement}
\bibfield{author}{\bibinfo{person}{Yvette Graham}, \bibinfo{person}{Timothy
  Baldwin}, \bibinfo{person}{Aaron Harwood}, \bibinfo{person}{Alistair Moffat},
  {and} \bibinfo{person}{Justin Zobel}.} \bibinfo{year}{2012}\natexlab{}.
\newblock \showarticletitle{Measurement of progress in machine translation}. In
  \bibinfo{booktitle}{\emph{Proceedings of the Australasian Language Technology
  Association Workshop 2012}}. \bibinfo{pages}{70--78}.
\newblock


\bibitem[\protect\citeauthoryear{Gu, Wang, Cho, and Li}{Gu
  et~al\mbox{.}}{2018}]%
        {gu2018search}
\bibfield{author}{\bibinfo{person}{Jiatao Gu}, \bibinfo{person}{Yong Wang},
  \bibinfo{person}{Kyunghyun Cho}, {and} \bibinfo{person}{Victor~OK Li}.}
  \bibinfo{year}{2018}\natexlab{}.
\newblock \showarticletitle{Search engine guided neural machine translation}.
  In \bibinfo{booktitle}{\emph{Thirty-Second AAAI Conference on Artificial
  Intelligence}}.
\newblock


\bibitem[\protect\citeauthoryear{Hao, Wang, Yang, Wang, Zhang, and Tu}{Hao
  et~al\mbox{.}}{2019}]%
        {hao-etal-2019-modeling}
\bibfield{author}{\bibinfo{person}{Jie Hao}, \bibinfo{person}{Xing Wang},
  \bibinfo{person}{Baosong Yang}, \bibinfo{person}{Longyue Wang},
  \bibinfo{person}{Jinfeng Zhang}, {and} \bibinfo{person}{Zhaopeng Tu}.}
  \bibinfo{year}{2019}\natexlab{}.
\newblock \showarticletitle{Modeling Recurrence for Transformer}. In
  \bibinfo{booktitle}{\emph{Proceedings of the 2019 Conference of the North
  {A}merican Chapter of the Association for Computational Linguistics: Human
  Language Technologies, Volume 1 (Long and Short Papers)}}.
  \bibinfo{publisher}{Association for Computational Linguistics},
  \bibinfo{address}{Minneapolis, Minnesota}, \bibinfo{pages}{1198--1207}.
\newblock
\urldef\tempurl%
\url{https://doi.org/10.18653/v1/N19-1122}
\showDOI{\tempurl}


\bibitem[\protect\citeauthoryear{Hassan, Aue, Chen, Chowdhary, Clark,
  Federmann, Huang, Junczys{-}Dowmunt, Lewis, Li, Liu, Liu, Luo, Menezes, Qin,
  Seide, Tan, Tian, Wu, Wu, Xia, Zhang, Zhang, and Zhou}{Hassan
  et~al\mbox{.}}{2018}]%
        {DBLP:journals/corr/abs-1803-05567}
\bibfield{author}{\bibinfo{person}{Hany Hassan}, \bibinfo{person}{Anthony Aue},
  \bibinfo{person}{Chang Chen}, \bibinfo{person}{Vishal Chowdhary},
  \bibinfo{person}{Jonathan Clark}, \bibinfo{person}{Christian Federmann},
  \bibinfo{person}{Xuedong Huang}, \bibinfo{person}{Marcin Junczys{-}Dowmunt},
  \bibinfo{person}{William Lewis}, \bibinfo{person}{Mu Li},
  \bibinfo{person}{Shujie Liu}, \bibinfo{person}{Tie{-}Yan Liu},
  \bibinfo{person}{Renqian Luo}, \bibinfo{person}{Arul Menezes},
  \bibinfo{person}{Tao Qin}, \bibinfo{person}{Frank Seide}, \bibinfo{person}{Xu
  Tan}, \bibinfo{person}{Fei Tian}, \bibinfo{person}{Lijun Wu},
  \bibinfo{person}{Shuangzhi Wu}, \bibinfo{person}{Yingce Xia},
  \bibinfo{person}{Dongdong Zhang}, \bibinfo{person}{Zhirui Zhang}, {and}
  \bibinfo{person}{Ming Zhou}.} \bibinfo{year}{2018}\natexlab{}.
\newblock \showarticletitle{Achieving Human Parity on Automatic Chinese to
  English News Translation}.
\newblock \bibinfo{journal}{\emph{CoRR}}  \bibinfo{volume}{abs/1803.05567}
  (\bibinfo{year}{2018}).
\newblock
\showeprint[arxiv]{1803.05567}
\urldef\tempurl%
\url{http://arxiv.org/abs/1803.05567}
\showURL{%
\tempurl}


\bibitem[\protect\citeauthoryear{Hazelwood, Bird, Brooks, Chintala, Diril,
  Dzhulgakov, Fawzy, Jia, Jia, Kalro, Law, Lee, Lu, Noordhuis, Smelyanskiy,
  Xiong, and Wang}{Hazelwood et~al\mbox{.}}{2018}]%
        {hazelwoodetal:fbleaner18}
\bibfield{author}{\bibinfo{person}{Kim Hazelwood}, \bibinfo{person}{Sarah
  Bird}, \bibinfo{person}{David Brooks}, \bibinfo{person}{Soumith Chintala},
  \bibinfo{person}{Utku Diril}, \bibinfo{person}{Dmytro Dzhulgakov},
  \bibinfo{person}{Mohamed Fawzy}, \bibinfo{person}{Bill Jia},
  \bibinfo{person}{Yangqing Jia}, \bibinfo{person}{Aditya Kalro},
  \bibinfo{person}{James Law}, \bibinfo{person}{Kevin Lee},
  \bibinfo{person}{Jason Lu}, \bibinfo{person}{Pieter Noordhuis},
  \bibinfo{person}{Misha Smelyanskiy}, \bibinfo{person}{Liang Xiong}, {and}
  \bibinfo{person}{Xiaodong Wang}.} \bibinfo{year}{2018}\natexlab{}.
\newblock \showarticletitle{Applied Machine Learning at {F}acebook: {A}
  Datacenter Infrastructure Perspective}. In \bibinfo{booktitle}{\emph{24th
  International Symposium on High-Performance Computer Architecture ({HPCA
  2018}), February 24-28, Vienna, Austria}}.
\newblock


\bibitem[\protect\citeauthoryear{Heigold, Varanasi, Neumann, and van
  Genabith}{Heigold et~al\mbox{.}}{2018}]%
        {DBLP:conf/amta/HeigoldVNG18}
\bibfield{author}{\bibinfo{person}{Georg Heigold}, \bibinfo{person}{Stalin
  Varanasi}, \bibinfo{person}{G{\"{u}}nter Neumann}, {and}
  \bibinfo{person}{Josef van Genabith}.} \bibinfo{year}{2018}\natexlab{}.
\newblock \showarticletitle{How Robust Are Character-Based Word Embeddings in
  Tagging and {MT} Against Wrod Scramlbing or Randdm Nouse?}. In
  \bibinfo{booktitle}{\emph{Proceedings of the 13th Conference of the
  Association for Machine Translation in the Americas, {AMTA} 2018, Boston, MA,
  USA, March 17-21, 2018 - Volume 1: Research Papers}}.
  \bibinfo{pages}{68--80}.
\newblock
\urldef\tempurl%
\url{https://aclanthology.info/papers/W18-1807/w18-1807}
\showURL{%
\tempurl}


\bibitem[\protect\citeauthoryear{Hunt and Szymanski}{Hunt and
  Szymanski}{1977}]%
        {hunt1977fast}
\bibfield{author}{\bibinfo{person}{James~W Hunt} {and}
  \bibinfo{person}{Thomas~G Szymanski}.} \bibinfo{year}{1977}\natexlab{}.
\newblock \showarticletitle{A fast algorithm for computing longest common
  subsequences}.
\newblock \bibinfo{journal}{\emph{Commun. ACM}} \bibinfo{volume}{20},
  \bibinfo{number}{5} (\bibinfo{year}{1977}), \bibinfo{pages}{350--353}.
\newblock


\bibitem[\protect\citeauthoryear{Jia and Harman}{Jia and Harman}{2011}]%
        {yjmh:analysis}
\bibfield{author}{\bibinfo{person}{Yue Jia} {and} \bibinfo{person}{Mark
  Harman}.} \bibinfo{year}{2011}\natexlab{}.
\newblock \showarticletitle{An Analysis and Survey of the Development of
  Mutation Testing}.
\newblock \bibinfo{journal}{\emph{{IEEE} Transactions on Software Engineering}}
  \bibinfo{volume}{37}, \bibinfo{number}{5} (\bibinfo{date}{September--October}
  \bibinfo{year}{2011}), \bibinfo{pages}{649 -- 678}.
\newblock


\bibitem[\protect\citeauthoryear{Jiang, Xiong, Zhang, Gao, and Chen}{Jiang
  et~al\mbox{.}}{2018}]%
        {Jiang:2018:SPR:3213846.3213871}
\bibfield{author}{\bibinfo{person}{Jiajun Jiang}, \bibinfo{person}{Yingfei
  Xiong}, \bibinfo{person}{Hongyu Zhang}, \bibinfo{person}{Qing Gao}, {and}
  \bibinfo{person}{Xiangqun Chen}.} \bibinfo{year}{2018}\natexlab{}.
\newblock \showarticletitle{Shaping Program Repair Space with Existing Patches
  and Similar Code}. In \bibinfo{booktitle}{\emph{Proceedings of the 27th ACM
  SIGSOFT International Symposium on Software Testing and Analysis}}
  \emph{(\bibinfo{series}{ISSTA 2018})}. \bibinfo{publisher}{ACM},
  \bibinfo{address}{New York, NY, USA}, \bibinfo{pages}{298--309}.
\newblock
\showISBNx{978-1-4503-5699-2}
\urldef\tempurl%
\url{https://doi.org/10.1145/3213846.3213871}
\showDOI{\tempurl}


\bibitem[\protect\citeauthoryear{Karpukhin, Levy, Eisenstein, and
  Ghazvininejad}{Karpukhin et~al\mbox{.}}{2019}]%
        {karpukhin2019training}
\bibfield{author}{\bibinfo{person}{Vladimir Karpukhin}, \bibinfo{person}{Omer
  Levy}, \bibinfo{person}{Jacob Eisenstein}, {and} \bibinfo{person}{Marjan
  Ghazvininejad}.} \bibinfo{year}{2019}\natexlab{}.
\newblock \showarticletitle{Training on Synthetic Noise Improves Robustness to
  Natural Noise in Machine Translation}.
\newblock \bibinfo{journal}{\emph{arXiv preprint arXiv:1902.01509}}
  (\bibinfo{year}{2019}).
\newblock


\bibitem[\protect\citeauthoryear{Khayrallah and Koehn}{Khayrallah and
  Koehn}{2018}]%
        {khayrallah2018impact}
\bibfield{author}{\bibinfo{person}{Huda Khayrallah} {and}
  \bibinfo{person}{Philipp Koehn}.} \bibinfo{year}{2018}\natexlab{}.
\newblock \showarticletitle{On the impact of various types of noise on neural
  machine translation}.
\newblock \bibinfo{journal}{\emph{arXiv preprint arXiv:1805.12282}}
  (\bibinfo{year}{2018}).
\newblock


\bibitem[\protect\citeauthoryear{Liu and Sun}{Liu and Sun}{2015}]%
        {liu2015contrastive}
\bibfield{author}{\bibinfo{person}{Yang Liu} {and} \bibinfo{person}{Maosong
  Sun}.} \bibinfo{year}{2015}\natexlab{}.
\newblock \showarticletitle{Contrastive unsupervised word alignment with
  non-local features}. In \bibinfo{booktitle}{\emph{Twenty-Ninth AAAI
  Conference on Artificial Intelligence}}.
\newblock


\bibitem[\protect\citeauthoryear{Manning, Surdeanu, Bauer, Finkel, Bethard, and
  McClosky}{Manning et~al\mbox{.}}{2014}]%
        {manningEtAl2014}
\bibfield{author}{\bibinfo{person}{Christopher~D. Manning},
  \bibinfo{person}{Mihai Surdeanu}, \bibinfo{person}{John Bauer},
  \bibinfo{person}{Jenny Finkel}, \bibinfo{person}{Steven~J. Bethard}, {and}
  \bibinfo{person}{David McClosky}.} \bibinfo{year}{2014}\natexlab{}.
\newblock \showarticletitle{The {Stanford} {CoreNLP} Natural Language
  Processing Toolkit}. In \bibinfo{booktitle}{\emph{Association for
  Computational Linguistics (ACL) System Demonstrations}}.
  \bibinfo{pages}{55--60}.
\newblock
\urldef\tempurl%
\url{http://www.aclweb.org/anthology/P/P14/P14-5010}
\showURL{%
\tempurl}


\bibitem[\protect\citeauthoryear{Mason}{Mason}{2017}]%
        {mason:lost}
\bibfield{author}{\bibinfo{person}{M.~Chris Mason}.}
  \bibinfo{year}{2017}\natexlab{}.
\newblock \bibinfo{title}{Strategic Insights: {L}ost in Translation}.
\newblock
\newblock
\urldef\tempurl%
\url{https://ssi.armywarcollege.edu/index.cfm/articles/Lost-In-Translation/2017/08/17}
\showURL{%
\tempurl}


\bibitem[\protect\citeauthoryear{Papineni, Roukos, Ward, and Zhu}{Papineni
  et~al\mbox{.}}{2002}]%
        {papineni2002bleu}
\bibfield{author}{\bibinfo{person}{Kishore Papineni}, \bibinfo{person}{Salim
  Roukos}, \bibinfo{person}{Todd Ward}, {and} \bibinfo{person}{Wei-Jing Zhu}.}
  \bibinfo{year}{2002}\natexlab{}.
\newblock \showarticletitle{{BLEU}: a method for automatic evaluation of
  machine translation}. In \bibinfo{booktitle}{\emph{Proceedings of the 40th
  annual meeting on association for computational linguistics}}. Association
  for Computational Linguistics, \bibinfo{pages}{311--318}.
\newblock


\bibitem[\protect\citeauthoryear{{Parmy Olson}}{{Parmy Olson}}{2018}]%
        {sexisttranslator}
\bibfield{author}{\bibinfo{person}{{Parmy Olson}}.}
  \bibinfo{year}{2018}\natexlab{}.
\newblock \bibinfo{title}{The Algorithm That Helped Google Translate Become
  Sexist}.
\newblock
  \bibinfo{howpublished}{\url{https://www.forbes.com/sites/parmyolson/2018/02/15/the-algorithm-that-helped-google-translate-become-sexist/##224101cb7daa}}.
\newblock


\bibitem[\protect\citeauthoryear{Pennington, Socher, and Manning}{Pennington
  et~al\mbox{.}}{2014}]%
        {pennington2014glove}
\bibfield{author}{\bibinfo{person}{Jeffrey Pennington},
  \bibinfo{person}{Richard Socher}, {and} \bibinfo{person}{Christopher
  Manning}.} \bibinfo{year}{2014}\natexlab{}.
\newblock \showarticletitle{Glove: {Global} vectors for word representation}.
  In \bibinfo{booktitle}{\emph{Proceedings of the 2014 conference on empirical
  methods in natural language processing (EMNLP)}}.
  \bibinfo{pages}{1532--1543}.
\newblock


\bibitem[\protect\citeauthoryear{Ribeiro, Singh, and Guestrin}{Ribeiro
  et~al\mbox{.}}{2018}]%
        {ribeiro2018semantically}
\bibfield{author}{\bibinfo{person}{Marco~Tulio Ribeiro},
  \bibinfo{person}{Sameer Singh}, {and} \bibinfo{person}{Carlos Guestrin}.}
  \bibinfo{year}{2018}\natexlab{}.
\newblock \showarticletitle{Semantically equivalent adversarial rules for
  debugging nlp models}. In \bibinfo{booktitle}{\emph{Proceedings of the 56th
  Annual Meeting of the Association for Computational Linguistics (Volume 1:
  Long Papers)}}. \bibinfo{pages}{856--865}.
\newblock


\bibitem[\protect\citeauthoryear{Ristad and Yianilos}{Ristad and
  Yianilos}{1998}]%
        {ristad1998learning}
\bibfield{author}{\bibinfo{person}{Eric~Sven Ristad} {and}
  \bibinfo{person}{Peter~N Yianilos}.} \bibinfo{year}{1998}\natexlab{}.
\newblock \showarticletitle{Learning string-edit distance}.
\newblock \bibinfo{journal}{\emph{IEEE Transactions on Pattern Analysis and
  Machine Intelligence}} \bibinfo{volume}{20}, \bibinfo{number}{5}
  (\bibinfo{year}{1998}), \bibinfo{pages}{522--532}.
\newblock


\bibitem[\protect\citeauthoryear{{Robert Parker, David Graff, Junbo Kong, Ke
  Chen, Kazuaki Maeda}}{{Robert Parker, David Graff, Junbo Kong, Ke Chen,
  Kazuaki Maeda}}{2011}]%
        {gigaword}
\bibfield{author}{\bibinfo{person}{{Robert Parker, David Graff, Junbo Kong, Ke
  Chen, Kazuaki Maeda}}.} \bibinfo{year}{2011}\natexlab{}.
\newblock \bibinfo{title}{{English Gigaword Fifth Edition}}.
\newblock
  \bibinfo{howpublished}{\url{https://catalog.ldc.upenn.edu/LDC2011T07}}.
\newblock


\bibitem[\protect\citeauthoryear{Saha, Lyu, Yoshida, and Prasad}{Saha
  et~al\mbox{.}}{2017}]%
        {Saha:2017:EEO:3155562.3155643}
\bibfield{author}{\bibinfo{person}{Ripon~K. Saha}, \bibinfo{person}{Yingjun
  Lyu}, \bibinfo{person}{Hiroaki Yoshida}, {and} \bibinfo{person}{Mukul~R.
  Prasad}.} \bibinfo{year}{2017}\natexlab{}.
\newblock \showarticletitle{ELIXIR: Effective Object Oriented Program Repair}.
  In \bibinfo{booktitle}{\emph{Proceedings of the 32Nd IEEE/ACM International
  Conference on Automated Software Engineering}} \emph{(\bibinfo{series}{ASE
  2017})}. \bibinfo{publisher}{IEEE Press}, \bibinfo{address}{Piscataway, NJ,
  USA}, \bibinfo{pages}{648--659}.
\newblock
\showISBNx{978-1-5386-2684-9}
\urldef\tempurl%
\url{http://dl.acm.org/citation.cfm?id=3155562.3155643}
\showURL{%
\tempurl}


\bibitem[\protect\citeauthoryear{{SpaCy}}{{SpaCy}}{2019}]%
        {SpaCy}
\bibfield{author}{\bibinfo{person}{{SpaCy}}.} \bibinfo{year}{2019}\natexlab{}.
\newblock \bibinfo{title}{{SpaCy}}.
\newblock \bibinfo{howpublished}{\url{https://spacy.io/}}.
\newblock


\bibitem[\protect\citeauthoryear{Sperber, Niehues, and Waibel}{Sperber
  et~al\mbox{.}}{2017}]%
        {sperber2017toward}
\bibfield{author}{\bibinfo{person}{Matthias Sperber}, \bibinfo{person}{Jan
  Niehues}, {and} \bibinfo{person}{Alex Waibel}.}
  \bibinfo{year}{2017}\natexlab{}.
\newblock \showarticletitle{Toward robust neural machine translation for noisy
  input sequences}. In \bibinfo{booktitle}{\emph{International Workshop on
  Spoken Language Translation (IWSLT)}}.
\newblock


\bibitem[\protect\citeauthoryear{Sun and Zhou}{Sun and Zhou}{2018}]%
        {sun2018metamorphic}
\bibfield{author}{\bibinfo{person}{Liqun Sun} {and} \bibinfo{person}{Zhi~Quan
  Zhou}.} \bibinfo{year}{2018}\natexlab{}.
\newblock \showarticletitle{Metamorphic testing for machine translations:
  MT4MT}. In \bibinfo{booktitle}{\emph{2018 25th Australasian Software
  Engineering Conference (ASWEC)}}. IEEE, \bibinfo{pages}{96--100}.
\newblock


\bibitem[\protect\citeauthoryear{Taylor, Marcus, and Santorini}{Taylor
  et~al\mbox{.}}{2003}]%
        {taylor2003penn}
\bibfield{author}{\bibinfo{person}{Ann Taylor}, \bibinfo{person}{Mitchell
  Marcus}, {and} \bibinfo{person}{Beatrice Santorini}.}
  \bibinfo{year}{2003}\natexlab{}.
\newblock \showarticletitle{The Penn treebank: an overview}.
\newblock In \bibinfo{booktitle}{\emph{Treebanks}}.
  \bibinfo{publisher}{Springer}, \bibinfo{pages}{5--22}.
\newblock


\bibitem[\protect\citeauthoryear{Vaswani, Bengio, Brevdo, Chollet, Gomez,
  Gouws, Jones, Kaiser, Kalchbrenner, Parmar, Sepassi, Shazeer, and
  Uszkoreit}{Vaswani et~al\mbox{.}}{2018}]%
        {tensor2tensor}
\bibfield{author}{\bibinfo{person}{Ashish Vaswani}, \bibinfo{person}{Samy
  Bengio}, \bibinfo{person}{Eugene Brevdo}, \bibinfo{person}{Francois Chollet},
  \bibinfo{person}{Aidan~N. Gomez}, \bibinfo{person}{Stephan Gouws},
  \bibinfo{person}{Llion Jones}, \bibinfo{person}{\L{}ukasz Kaiser},
  \bibinfo{person}{Nal Kalchbrenner}, \bibinfo{person}{Niki Parmar},
  \bibinfo{person}{Ryan Sepassi}, \bibinfo{person}{Noam Shazeer}, {and}
  \bibinfo{person}{Jakob Uszkoreit}.} \bibinfo{year}{2018}\natexlab{}.
\newblock \showarticletitle{Tensor2Tensor for Neural Machine Translation}.
\newblock \bibinfo{journal}{\emph{CoRR}}  \bibinfo{volume}{abs/1803.07416}
  (\bibinfo{year}{2018}).
\newblock
\urldef\tempurl%
\url{http://arxiv.org/abs/1803.07416}
\showURL{%
\tempurl}


\bibitem[\protect\citeauthoryear{Vaswani, Shazeer, Parmar, Uszkoreit, Jones,
  Gomez, Kaiser, and Polosukhin}{Vaswani et~al\mbox{.}}{2017}]%
        {vaswani2017attention}
\bibfield{author}{\bibinfo{person}{Ashish Vaswani}, \bibinfo{person}{Noam
  Shazeer}, \bibinfo{person}{Niki Parmar}, \bibinfo{person}{Jakob Uszkoreit},
  \bibinfo{person}{Llion Jones}, \bibinfo{person}{Aidan~N Gomez},
  \bibinfo{person}{{\L}ukasz Kaiser}, {and} \bibinfo{person}{Illia
  Polosukhin}.} \bibinfo{year}{2017}\natexlab{}.
\newblock \showarticletitle{Attention is all you need}. In
  \bibinfo{booktitle}{\emph{Proceedings of the 31st International Conference on
  Neural Information Processing Systems}}. Curran Associates Inc.,
  \bibinfo{pages}{6000--6010}.
\newblock


\bibitem[\protect\citeauthoryear{{VoiceBoxer}}{{VoiceBoxer}}{2016}]%
        {chinese01}
\bibfield{author}{\bibinfo{person}{{VoiceBoxer}}.}
  \bibinfo{year}{2016}\natexlab{}.
\newblock \bibinfo{title}{WHAT ABOUT ENGLISH IN CHINA?}
\newblock
  \bibinfo{howpublished}{\url{http://voiceboxer.com/english-in-china/}}.
\newblock


\bibitem[\protect\citeauthoryear{Wei and Su}{Wei and Su}{2012}]%
        {wei2012statistics}
\bibfield{author}{\bibinfo{person}{Rining Wei} {and} \bibinfo{person}{Jinzhi
  Su}.} \bibinfo{year}{2012}\natexlab{}.
\newblock \showarticletitle{The statistics of English in China: An analysis of
  the best available data from government sources}.
\newblock \bibinfo{journal}{\emph{English Today}} \bibinfo{volume}{28},
  \bibinfo{number}{3} (\bibinfo{year}{2012}), \bibinfo{pages}{10--14}.
\newblock


\bibitem[\protect\citeauthoryear{{Wikipedia}}{{Wikipedia}}{2014}]%
        {wikipedia}
\bibfield{author}{\bibinfo{person}{{Wikipedia}}.}
  \bibinfo{year}{2014}\natexlab{}.
\newblock \bibinfo{title}{{Wikipedia}}.
\newblock \bibinfo{howpublished}{\url{https://dumps.wikimedia.org/}}.
\newblock


\bibitem[\protect\citeauthoryear{{WMT}}{{WMT}}{2018}]%
        {news}
\bibfield{author}{\bibinfo{person}{{WMT}}.} \bibinfo{year}{2018}\natexlab{}.
\newblock \bibinfo{title}{{News-Commentary}}.
\newblock
  \bibinfo{howpublished}{\url{http://data.statmt.org/wmt18/translation-task/}}.
\newblock


\bibitem[\protect\citeauthoryear{Xin and Reiss}{Xin and Reiss}{2017}]%
        {Xin:2017:LSC:3155562.3155644}
\bibfield{author}{\bibinfo{person}{Qi Xin} {and} \bibinfo{person}{Steven~P.
  Reiss}.} \bibinfo{year}{2017}\natexlab{}.
\newblock \showarticletitle{Leveraging Syntax-related Code for Automated
  Program Repair}. In \bibinfo{booktitle}{\emph{Proceedings of the 32Nd
  IEEE/ACM International Conference on Automated Software Engineering}}
  \emph{(\bibinfo{series}{ASE 2017})}. \bibinfo{publisher}{IEEE Press},
  \bibinfo{address}{Piscataway, NJ, USA}, \bibinfo{pages}{660--670}.
\newblock
\showISBNx{978-1-5386-2684-9}
\urldef\tempurl%
\url{http://dl.acm.org/citation.cfm?id=3155562.3155644}
\showURL{%
\tempurl}


\bibitem[\protect\citeauthoryear{Zhang, Chen, Hao, Xiong, Xie, Zhang, and
  Mei}{Zhang et~al\mbox{.}}{2014}]%
        {zhang2014search}
\bibfield{author}{\bibinfo{person}{Jie Zhang}, \bibinfo{person}{Junjie Chen},
  \bibinfo{person}{Dan Hao}, \bibinfo{person}{Yingfei Xiong},
  \bibinfo{person}{Bing Xie}, \bibinfo{person}{Lu Zhang}, {and}
  \bibinfo{person}{Hong Mei}.} \bibinfo{year}{2014}\natexlab{}.
\newblock \showarticletitle{Search-based inference of polynomial metamorphic
  relations}. In \bibinfo{booktitle}{\emph{Proceedings of the 29th ACM/IEEE
  international conference on Automated software engineering}}. ACM,
  \bibinfo{pages}{701--712}.
\newblock


\bibitem[\protect\citeauthoryear{Zhang, Utiyama, Sumita, Neubig, and
  Nakamura}{Zhang et~al\mbox{.}}{2018}]%
        {zhang2018guiding}
\bibfield{author}{\bibinfo{person}{Jingyi Zhang}, \bibinfo{person}{Masao
  Utiyama}, \bibinfo{person}{Eiichro Sumita}, \bibinfo{person}{Graham Neubig},
  {and} \bibinfo{person}{Satoshi Nakamura}.} \bibinfo{year}{2018}\natexlab{}.
\newblock \showarticletitle{Guiding neural machine translation with retrieved
  translation pieces}.
\newblock \bibinfo{journal}{\emph{arXiv preprint arXiv:1804.02559}}
  (\bibinfo{year}{2018}).
\newblock


\bibitem[\protect\citeauthoryear{Zhang, Harman, Ma, and Liu}{Zhang
  et~al\mbox{.}}{2019}]%
        {zhang2019machine}
\bibfield{author}{\bibinfo{person}{Jie~M Zhang}, \bibinfo{person}{Mark Harman},
  \bibinfo{person}{Lei Ma}, {and} \bibinfo{person}{Yang Liu}.}
  \bibinfo{year}{2019}\natexlab{}.
\newblock \showarticletitle{Machine Learning Testing: Survey, Landscapes and
  Horizons}.
\newblock \bibinfo{journal}{\emph{arXiv preprint arXiv:1906.10742}}
  (\bibinfo{year}{2019}).
\newblock


\bibitem[\protect\citeauthoryear{Zhang, Jin, and Zhou}{Zhang
  et~al\mbox{.}}{2010}]%
        {Zhang2010}
\bibfield{author}{\bibinfo{person}{Yin Zhang}, \bibinfo{person}{Rong Jin},
  {and} \bibinfo{person}{Zhi-Hua Zhou}.} \bibinfo{year}{2010}\natexlab{}.
\newblock \showarticletitle{Understanding bag-of-words model: a statistical
  framework}.
\newblock \bibinfo{journal}{\emph{International Journal of Machine Learning and
  Cybernetics}} \bibinfo{volume}{1}, \bibinfo{number}{1} (\bibinfo{date}{01
  Dec} \bibinfo{year}{2010}), \bibinfo{pages}{43--52}.
\newblock
\showISSN{1868-808X}
\urldef\tempurl%
\url{https://doi.org/10.1007/s13042-010-0001-0}
\showDOI{\tempurl}


\bibitem[\protect\citeauthoryear{Zhao, Dua, and Singh}{Zhao
  et~al\mbox{.}}{2017}]%
        {DBLP:journals/corr/abs-1710-11342}
\bibfield{author}{\bibinfo{person}{Zhengli Zhao}, \bibinfo{person}{Dheeru Dua},
  {and} \bibinfo{person}{Sameer Singh}.} \bibinfo{year}{2017}\natexlab{}.
\newblock \showarticletitle{Generating Natural Adversarial Examples}.
\newblock \bibinfo{journal}{\emph{CoRR}}  \bibinfo{volume}{abs/1710.11342}
  (\bibinfo{year}{2017}).
\newblock
\showeprint[arxiv]{1710.11342}
\urldef\tempurl%
\url{http://arxiv.org/abs/1710.11342}
\showURL{%
\tempurl}


\bibitem[\protect\citeauthoryear{Ziemski, Junczys-Dowmunt, and
  Pouliquen}{Ziemski et~al\mbox{.}}{2016}]%
        {UN}
\bibfield{author}{\bibinfo{person}{Micha{\l} Ziemski}, \bibinfo{person}{Marcin
  Junczys-Dowmunt}, {and} \bibinfo{person}{Bruno Pouliquen}.}
  \bibinfo{year}{2016}\natexlab{}.
\newblock \showarticletitle{The united nations parallel corpus v1. 0}. In
  \bibinfo{booktitle}{\emph{Proceedings of the Tenth International Conference
  on Language Resources and Evaluation (LREC 2016)}}.
  \bibinfo{pages}{3530--3534}.
\newblock


\end{thebibliography}
\bibliographystyle{ACM-Reference-Format}
\vspace{12pt}
\end{document}